\documentclass[preprint,12pt]{elsarticle}
\usepackage{amsmath}
\usepackage{lineno}
\usepackage{array}
\usepackage{multirow}
\usepackage{epstopdf}
\modulolinenumbers[5]
\journal{Int. J. Rock Mech. Min. Sci.}

\begin{document}
	
	
	\begin{titlepage}
		\clearpage\thispagestyle{empty}
		\noindent
		\hrulefill
		\begin{figure}[h!]
			\centering
			\includegraphics[width=2 in]{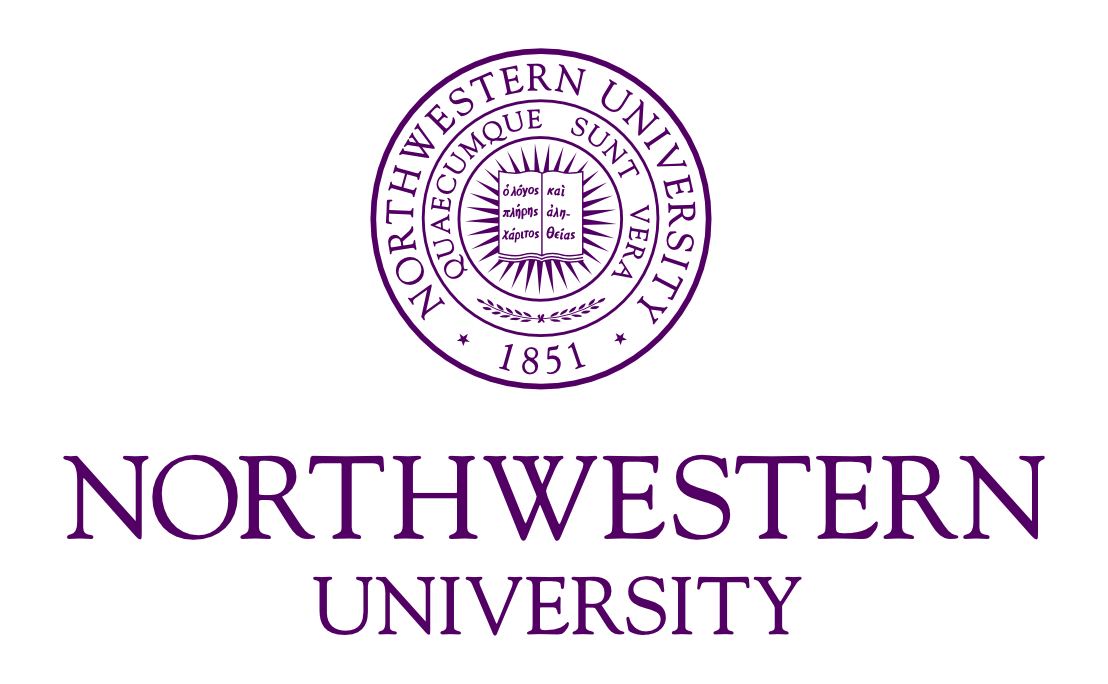}
		\end{figure}
		\begin{center}
			{
				{
					{\bf Center for Sustainable Engineering of Geological and Infrastructure Materials} \\ [0.1in]
					Department of Civil and Environmental Engineering \\ [0.1in]
					McCormick School of Engineering and Applied Science \\ [0.1in]
					Evanston, Illinois 60208, USA
				}
			}
		\end{center} 
		\hrulefill \\ \vskip 2mm
		\begin{center}
			{\large {\bf ELASTIC, STRENGTH, AND FRACTURE PROPERTIES OF MARCELLUS SHALE
				}}\\[0.5in]
				{\large {\sc Zhefei Jin, Weixin Li, Congrui Jin, James Hambleton, Gianluca Cusatis}}\\[0.75in]
				{\sf \bf SEGIM INTERNAL REPORT No. 17-12/478E}\\[0.75in]
			\end{center}
			\vskip 5mm
			\noindent {\footnotesize {{\em Submitted to Int. J. Rock Mech. Min. Sci. \hfill December 2017} }}
		\end{titlepage}
		
		\newpage
		\clearpage \pagestyle{plain} \setcounter{page}{1}
        
\begin{frontmatter}

\title{Elastic, strength, and fracture properties of Marcellus shale}


\author[mymainaddress]{Zhefei Jin}\ead{ZhefeiJin2015@u.northwestern.edu}
\author[mysecondaryaddress]{Weixin Li}\ead{w.li@u.northwestern.edu}
\author[mythirdaddress]{Congrui Jin}\ead{cjin@binghamton.edu}
\author[mymainaddress]{James Hambleton}\ead{jphambleton@northwestern.edu}
\author[mymainaddress,mysecondaryaddress]{ Gianluca Cusatis\corref{mycorrespondingauthor}}
\cortext[mycorrespondingauthor]{Corresponding author}
\ead{g-cusatis@northwestern.edu}

\address[mymainaddress]{Department of Civil and Environmental Engineering, Northwestern University, Evanston, IL 60208, U.S.A}
\address[mysecondaryaddress]{Theoretical and Applied Mechanics, Northwestern University, Evanston, IL 60208, U.S.A}
\address[mythirdaddress]{Department of Mechanical Engineering, State University of New York at Binghamton, Binghamton, NY 13902, U.S.A}

\begin{abstract}
Shale, a fine-grained sedimentary rock, is the key source rock for many of the world's most important oil and natural gas deposits. A deep understanding of the mechanical properties of shale is of vital importance in various geotechnical applications, including oil and gas exploitation. In this work, deformability, strength, and fracturing properties of Marcellus shale were investigated through an experimental study. Firstly, uniaxial compression, direct tension, and Brazilian tests were performed on the Marcellus shale specimens in various bedding plane orientations with respect to loading directions to measure the static mechanical properties and their anisotropy. Furthermore, the deformability of Marcellus shale was also studied through seismic velocity measurements for comparison with the static measurements. The experimental results revealed that the transversely isotropic model is applicable for describing the elastic behaviors of Marcellus shale in pure tension and compression. The elastic properties measured from these two experiments, however, were not exactly the same. Strength results showed that differences exist between splitting (Brazilian) and direct tensile strengths, both of which varied with bedding plane orientations and loading directions and were associated with different failure modes. Finally, a series of three-point-bending tests were conducted on specimens of increasing size in three different principal notch orientations to investigate the fracture properties of the material. It was found that there exists a significant size effect on the fracture properties calculated from the measured peak loads and by using the Linear Elastic Fracture Mechanics (LEFM) theory. The fracture properties can be uniquely identified, however, by using Ba\v zant's Size Effect Law and they were found to be anisotropic. 	
\end{abstract}

\end{frontmatter}

\section{Introduction}	
In recent years, shale has been widely studied due to its role in various subdomains of geomechanics, especially in relation to the oil and gas industries. The geomechanics of shale reservoirs is a subject of particurlar interest attracting sustained interest from researchers and practitioners alike. A thorough experimental campaign for shale characterization is needed to order to interpret its material properties, and to provide fundamental insight into industrial processes and applications such as hydraulic fracturing, subsurface carbon dioxide sequestration, and nuclear waste disposal. 

Shale rocks, sometimes referred to as mudstones, are composed of extremely fine-grained particles. Their mineralogical composition exhibits a wide range, including clay, quartz, feldspar, and other minerals \cite{li2016integrated}. The heterogeneities exist at different length scales ranging from microscopic level to an entire rock mass, which leads to a large variety of macroscopic behaviors \cite{li2016integrated,li2017multiscale}. As a result, the experimentally determined properties are not single-valued, well-defined parameters for a given type of shale. In addition, shale is generally regarded as a transversely isotropic material \cite{jones1981ultrasonic}, and the measured properties vary with the orientations in which the sample is cored, as well as the loading directions. This anisotropy can be attributed to several factors, including preferred orientation of clay minerals \cite{jones1981ultrasonic, johnston1995seismic,hornby1998experimental}, presence of fluid-filled microcracks \cite{hornby1994anisotropic}, stress state and history \cite{holt1991influence,sayers1999stress}, and kerogen content \cite{vernik1996elastic}. Overall, it is of vital importance to build a comprehensive database of shale characteristics taking into account its heterogeneity and anisotropy, in order to calibrate any analytical theories or computational frameworks. 

A volume of work has been completed on the mechanical characterization of shale. In general, deformability, strength, and fracturability are the focus of many experimental studies as they are fundamental characteristics necessary for any field or laboratory investigations. To explore the mechanical properties comprehensively, a variety of experimental methodologies were proposed. 

Uniaxial tests have been used widely to measure the strength of shale in both tension and compression. For measurements of tensile strength, indirect methods such as the Brazilian test have commonly been used to avoid difficulties associated with performing direct uniaxial tension tests.  The Brazilian test, also known as splitting tension test, was performed on 
various types of shale-like rocks, such as Woodford, Mancos, Boryeong, and Montney shale. The test results strongly depended on the loading directions, suggesting significant anisotropy of the inferred strength, referred to as the Brazilian Tensile Strength (BTS) \cite{cho2012deformation,keneti2010investigation,vervoort2014failure,sierra2010woodford}. Keneti and Wong \cite{keneti2010investigation} and Sierra et al. \cite{sierra2010woodford} measured the BTS of Montney and Woodford shale, respectively, in two directions: one in which the loading direction is parallel to the plane of isotropy, which coincides with the bedding plane (see Section \ref{Sec: 2P1}), and another in which the loading direction is perpendicular to that plane. It was found that the values of BTS obtained for these shales were larger when the loading direction was perpendicular to the isotropic plane of the materials. Cho et al. \cite{cho2012deformation, vervoort2014failure} measured the BTS of Boryeong shale with different inclination angles between the isotropic plane of the material and the orientation of the applied loads (or loading direction), denoted as $\beta$. The results revealed that with an increase of the inclination angle, the measured BTS values almost kept constant at first, and then decreased. It was also found that different failure patterns associated with the different BTS values corresponding to different inclination angles may occur. For Boryeong shale, the failure occurred along the loading direction for  $0^\circ\leq\beta\leq15^\circ$, and it started to fail along the bedding plane for $\beta\geq30^\circ$ \cite{cho2012deformation, vervoort2014failure}. For Mancos shale, the failure path was straight for $60^\circ\leq\beta\leq90^\circ$ and $\beta = 0^\circ$, or followed a zig-zag pattern for $15^\circ\leq\beta\leq45^\circ$ \cite{simpson2014failure}. Uniaxial Compressive Strength (UCS) is another important rock characteristic which is usually measured through the uniaxial compression test. It was found that the UCS of shale is strongly related to the mineral composition and structure. For instance, Jizba \cite{jizba1992mechanical} found that the UCS increases with shale clay content. On account of this, the UCS measurements scatter significantly among different types of shale: the mean value of the UCS for Boryeong shale is about $90$ MPa, and that for Eagle Ford shale is about $3$ MPa. Similar to BTS, the measured UCS values of Boryeong and Eagle Ford shale also showed a distinct anisotropy. The ratio of maximum to minimum UCS measurements for Boryeong shale was about $2.6$, and that for Eagle Ford shale was about $11$. The maximum UCS was found to occur when the loading direction was parallel to bedding, whereas the minimum occurred when the samples failed along bedding \cite{cho2012deformation,hsu2002characterization}.

The deformability of shale was explored through both static and dynamic methods. Uniaxial and triaxial compression tests were widely used to measure the compressive moduli and elastic constants of shale. Lee et al. \cite{lee2015interaction} performed uniaxial compression tests on Marcellus shale samples, and the Young's modulus of the specimens with loading direction parallel to bedding measured $8.9$ GPa. The same test was conducted on Boryeong shale \cite{cho2012deformation} to investigate the deformation anisotropy. The results showed that the compressive modulus varies from $19$ to $39$ GPa as the angle between the loading direction and the bedding plane varies from $0$ to $90^\circ$. Similarly, the deformational properties obtained through triaxial compression tests were found to be directionally dependent  \cite{niandou1997laboratory,smith1969fracture,salager2013constitutive}.  It was found that the  elastic moduli of the specimens obtained with the loading orientation perpendicular to bedding was larger than that of the specimens with loading parallel to bedding. Apart from deformational anisotropy, pressure dependency of the deformability was also observed, as determined through triaxial tests. For instance, Lora et al. \cite{lora2016geomechanical} found that Marcellus shale specimens became stiffer as the confining pressure increased, and the mechanical response became more linear. The degree of deformational anisotropy also reduced as the confining pressure increased. Seismic velocities, from which elastic properties can be extracted, were also measured as a means of characterizing shale deformability. As in static tests, significant anisotropy of the seismic velocity was frequently observed \cite{wang2001seismic,vernik1997velocity,banik1984velocity,vernik1992ultrasonic,kim2012anisotropy}. The existing database of seismic velocity measurements supports the assumption that shale rocks can be considered as transversely isotropic media \cite{jones1981ultrasonic,johnston1995seismic}. Based on the transversely isotropic model, five elastic constants can be determined (see Section \ref{Sec: 2P3}) and related to the longitudinal and transverse wave (P-wave and S-wave ) velocities \cite{white1983measured,white1953velocity,sone2013mechanical}. Although seismic velocity measurement is an inexpensive and quick way to determine rock deformational properties, it is not enough in a rock engineering project \cite{b1987general}, the reason being that there exist non-negligible differences between statically and dynamically determined elastic properties \cite{homand1993characterization,cheng1981dynamic,christaras1994determination}. For example, static measurements are liable to be more affected by the presence of cracks, cavities and a non-linear elastic response than the dynamics measurements. This leads the statically determined Young's modulus to be comparatively low \cite{ide1936comparison,eissa1988relation}.    

Shale fracturability describes the material's resistance to fracture, and is usually quantified by static fracture toughness based on the theory of Linear Elastic Fracture Mechanics (LEFM). Mode I fracture toughness, denoted $K_{Ic}$, is the most widely used fracture characteristic, and is measured when a sample containing a notch or traction-free crack is tested under a normal tensile stress perpendicular to the crack. The $K_{Ic}$ data are common and were determined in laboratory tests for different types of shale rocks, including Mancos \cite{chandler2016fracture}, Marcellus \cite{lee2015interaction}, and Woodford \cite{sierra2010woodford}. Among these shale rocks, fracture toughness tends to exhibit a large scatter, ranging from 0.1 to 1.5 MPa$\sqrt{\text{m}}$. Even for the same type of rock, with samples from the same core, a large variation of $K_{Ic}$ was frequently observed \cite{wang2017experimental}. Indeed, the laboratory determined fracture properties of rocks and other geomaterials were strongly related to the testing setup, especially the specimen geometry and sample size \cite{chong1989size,chong1984mechanics,ingraffea1984short,khan2000effect,ayatollahi2014size,bavzant1991identification,ba1990determination}. Despite the abundance of fracture data, experimental studies of shale fracture properties considering the specimen size and geometry dependency are limited.  

In this work, a series of experiments were performed on Marcellus shale samples to obtain a comprehensive experimental database of shale strength, deformability, and fracturability. 

\section{Theoretical background} 
\subsection{Compliance matrix for transversely isotropic shale}\label{Sec: 2P1}
\begin{figure}
    \begin{center}
         \includegraphics[scale=0.47, trim= 12mm 78mm 70mm 20mm, clip]{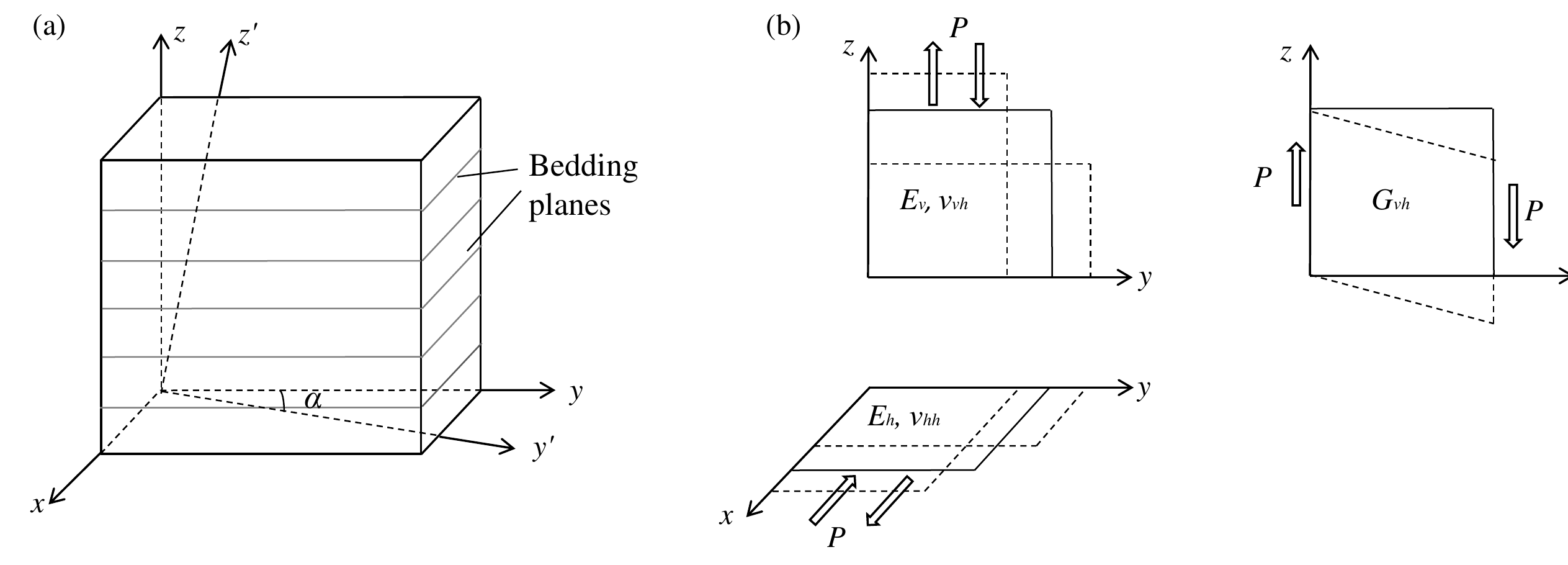}\\
        \caption{(a) Coordinate system. (b) Schematic diagrams for the meaning of the five elastic constants. }
        \label{2.10}
        \end{center}
\end{figure}

At the macroscopic level, shale consists of a layered sedimentary rock, and is often considered as a transversely isotropic continuum. Under this assumption, the material properties are symmetric about an axis normal to a plane of isotropy which coincides with the bedding plane. The coordinate system for this study on the elasticity of Marcellus shale is illustrated in Fig.~\ref{2.10}a. The horizontal plane ($x\text{-}y$ plane) represents the plane of isotropy, and the mechanical properties are same in all directions within this plane. Normal to the $x\text{-}y$ plane, the $z$-axis represents the axis of symmetry. With the coordinate system introduced above, the strain and stress in the elastic regime can be related by Hooke's law, by using Voigt notation:
\begin{equation}\label{14}
\left\{\begin{array}{l}
\varepsilon_{x}\\\varepsilon_{y}\\\varepsilon_{z}\\\varepsilon_{xy}\\\varepsilon_{xz}\\\varepsilon_{yz}\end{array}\right\}   
=\left[\begin{array}{rrrrrr}
\frac{1}{E_{h}}&-\frac{\nu_{hh}}{E_{h}}&-\frac{\nu_{vh}}{E_{v}}&&&\\
-\frac{\nu_{hh}}{E_{h}}&\frac{1}{E_{h}}&-\frac{\nu_{hh}}{E_{v}}&&&\\
-\frac{\nu_{vh}}{E_{v}}&-\frac{\nu_{vh}}{E_{v}} &\frac{1}{E_{v}}&&&\\
&&&\frac{2(1+\nu_{hh})}{E_{h}}&&\\
&&&&\frac{1}{G_{vh}}&\\
&&&&&\frac{1}{G_{vh}}\\\end{array}\right]
\left\{\begin{array}{l}
\sigma_{x}\\\sigma_{y}\\\sigma_{z}\\\sigma_{xy}\\\sigma_{xz}\\\sigma_{yz}\end{array}\right\}
\end{equation}
where $E_h$ and $E_v$ are the Young's modulus in the horizontal ($x\text{-}y$ plane, or bedding plane) and vertical (normal to $x\text{-}y$ plane) directions respectively; $\nu_{hh}$ is the Poisson's ratio characterizing the Poisson effect in the horizontal plane, while $\nu_{vh}$ represents the Poisson’s ratio in the vertical plane ($x\text{-}z$ or $y\text{-}z$ plane); $G_{vh}$ represents the shear modulus in the vertical plane. The schematic diagrams for these five independent elastic constants are shown in Fig.~\ref{2.10}b. 

A new coordinate system, $x'\text{-}y'\text{-}z'$, is defined by rotating the original coordinate system, $x\text{-}y\text{-}z$, by angle $\alpha$ in a clockwise direction about the $x$-axis (see Fig.~\ref{2.10}a). The generalized Hooke's law in this coordinate system can be written as
\begin{equation}\label{15}
{\varepsilon}'_{ij} =S'_{ijkl}{\sigma}'_{kl},
\end{equation}
where $S'_{ijkl}$ is the fourth-order compliance tensor defined with respect to the new coordinate system. By using Voigt notation, the above equation can be written as
\begin{equation}\label{16}
\left\{\begin{array}{c}
\varepsilon_{x'}\\\varepsilon_{y'}\\\varepsilon_{z'}\\\varepsilon_{x'y'}\\\varepsilon_{x'z'}\\\varepsilon_{y'z'}\end{array}\right\}
=\left[\begin{array}{cccccc}
S'_{11}&S'_{12}&S'_{13}&S'_{14}&S'_{15}&S'_{16}\\
S'_{21}&S'_{22}&S'_{23}&S'_{24}&S'_{25}&S'_{26}\\
S'_{31}&S'_{32}&S'_{33}&S'_{34}&S'_{35}&S'_{36}\\
S'_{41}&S'_{42}&S'_{43}&S'_{44}&S'_{45}&S'_{46}\\
S'_{51}&S'_{52}&S'_{53}&S'_{54}&S'_{55}&S'_{56}\\
S'_{61}&S'_{62}&S'_{63}&S'_{64}&S'_{65}&S'_{66}\\\end{array}\right]
\left\{\begin{array}{c}
\sigma_{x'}\\\sigma_{y'}\\\sigma_{z'}\\\sigma_{x'y'}\\\sigma_{x'z'}\\\sigma_{y'z'}\end{array}\right\}   
\end{equation}

By taking the coordinate transformation of a fourth-order tensor, the components of the compliance tensor in the new coordinate system $S'_{ijkl}$ can be related to the ones in the original system $S_{ijkl}$. Accordingly, the component $S'_{33}$ can be determined, and it is related to the Young's modulus associated with the $z'$-axis in the new coordinate system, $E'_v$, as follows:
\begin{equation}\label{Eq: E33}
S'_{33}=\frac{\varepsilon'_z}{\sigma'_z}=\frac{1}{E'_v}=\frac{\sin^4\alpha}{E_h}+\frac{\cos^4\alpha}{E_v}+\frac{\sin^22\alpha}{4}(\frac{1}{G_{vh}}-\frac{2\nu_{vh}}{E_v})
\end{equation} 

\subsection{Determination of elastic constants}\label{Sec: EC}

\begin{figure}
	\begin{center}
		\includegraphics[width = 0.35\textwidth]{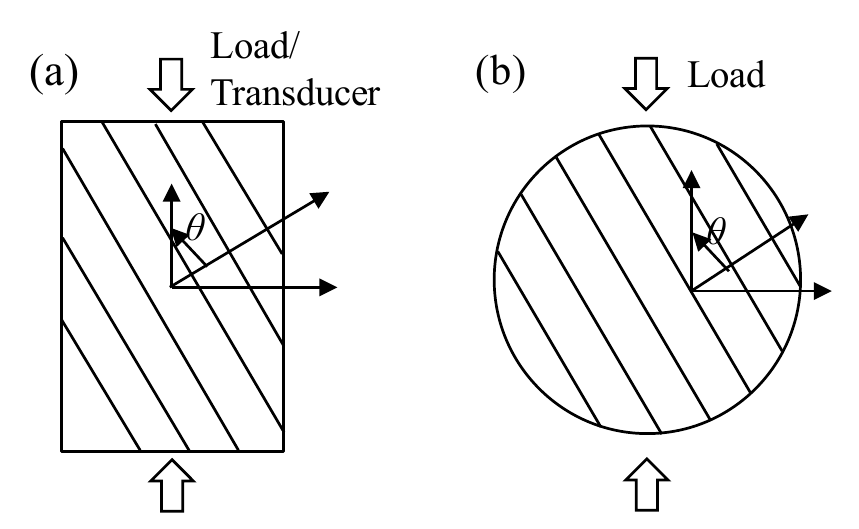}\\
		\caption{Definition of anisotropy angle for (a) cubic/cuboid and (b) disc specimens.}
		\label{Fig:Anisotropy}
	\end{center}
\end{figure}

\begin{figure}
	\begin{center}
		\includegraphics[scale=0.65, trim= 52mm 65mm 0mm 45mm, clip]{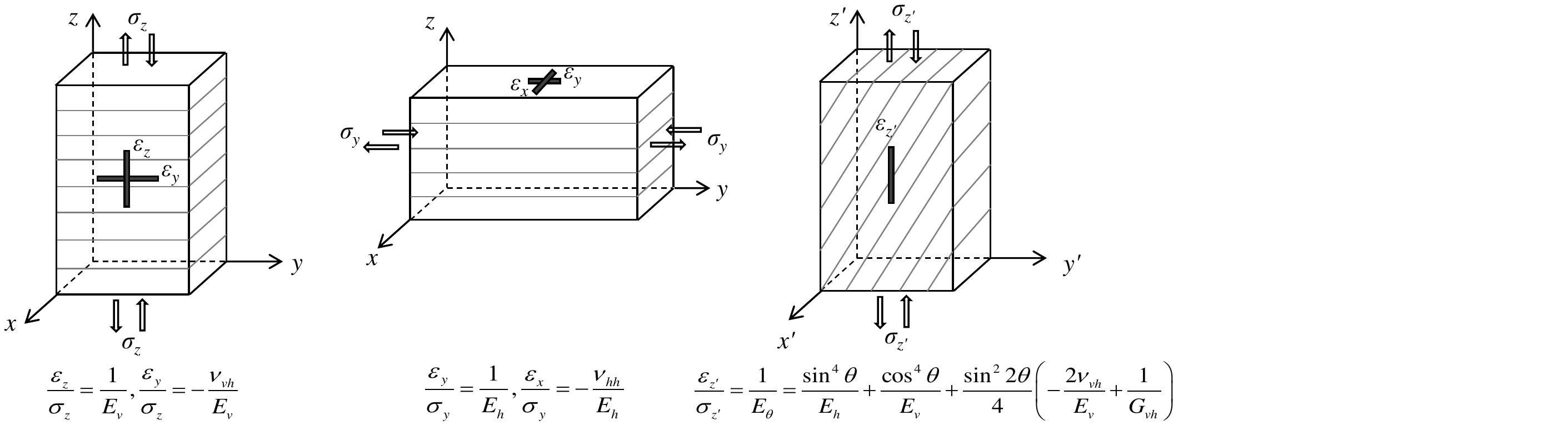}\\
		\caption{Specimens and corresponding equation for determining elastic constants}
		\label{2.12}
	\end{center}
\end{figure}

The method to determine experimentally the five elastic constants has been devised in previous studies \cite{amadei1996importance,pinto1970deformability,hakala2007estimating,cho2012deformation}. In this method, uniaxial tension or compression tests are conducted. At least three specimens, two with anisotropy angles $\theta=0^{\circ}$ and $90^{\circ}$, and one with $\theta$ between $0^{\circ}$ and $90^{\circ}$, are considered. Here, the anisotropy angle $\theta$ is defined as the angle between the loading direction and normal vector to the bedding plane, as shown in Fig~\ref{Fig:Anisotropy}. The independent equations relevant to each type of specimen are reported in Fig.~\ref{2.12}. For specimens with $\theta= 0^\circ$ and $90^\circ$ under vertical loading, strains are measured in two orthogonal directions in the vertical plane, and correspondingly two independent equations, Eqs. \ref{23} and \ref{231}, are obtained. For specimens with $0^\circ\leq\theta\leq90^\circ$, only axial strain is measured, and correspondingly Eq. \ref{232} is obtained. In these equations, stress ($\sigma$), strain ($\varepsilon$) and anisotropy angle ($\theta$) are the measured values. 

\begin{equation}\label{23}
\frac{\varepsilon_{z}}{\sigma_{z}}=\frac{1}{E_{v}}, \frac{\varepsilon_{y}}{\sigma_{z}}=-\frac{\nu_{vh}}{E_{v}}
\end{equation}

\begin{equation}\label{231}
\frac{\varepsilon_{y}}{\sigma_{y}}=\frac{1}{E_{h}}, \frac{\varepsilon_{x}}{\sigma_{y}}=-\frac{\nu_{hh}}{E_{h}}
\end{equation}	

\begin{equation}\label{232}
\frac{\varepsilon_{z'}}{\sigma_{z'}}=\frac{\sin^4\theta}{E_h}+\frac{\cos^4\theta}{E_v}+\frac{\sin^22\theta}{4}(-\frac{2\nu_{vh}}{E_v}+\frac{1}{G_{vh}})
\end{equation}  

In this work, uniaxial compression and direct tension tests were performed on the specimens of anisotropy angle $0^\circ, 30^\circ, 45^\circ, 60^\circ$ and $90^\circ$. As a consequence, seven independent equations were obtained for the determination of the elastic constants under tension or compression: tests on the specimens with $\theta= 0^\circ$ and $90^\circ$ complete four of them, and tests on the specimens with $\theta= 30^\circ, 45^\circ$, and $60^\circ$ complete three of them. The five elastic constants were estimated out of the seven equations by utilizing the least-squares method. By combining Eqs. \ref{23}-\ref{232}, one can rewrite the overdetermined equations in a matrix form with the unknowns on the left-hand side representing the elastic constants, while the right-hand side represents experimental measurements: 
\begin{equation}\label{Eq: over}
\footnotesize
\setlength{\arraycolsep}{5
pt} 
\left[
\begin{array}{ccccc}
1&0&0&0&0\\
0&-1&0&0&0\\
0&0&1&0&0\\
0&0&0&-1&0\\
\cos^4\theta_1&-2\sin^22\theta_1/4&\sin^4\theta_1&0&\sin^22\theta_1/4\\
\cos^4\theta_2&-2\sin^22\theta_2/4&\sin^4\theta_2&0&\sin^22\theta_2/4\\
\cos^4\theta_3&-2\sin^22\theta_3/4&\sin^4\theta_3&0&\sin^22\theta_3/4\\\end{array}\right]
\left\{\begin{array}{c}
\frac{1}{E_v}\\\frac{\nu_{vh}}{E_v}\\\frac{1}{E_h}\\\frac{\nu_{hh}}{E_h}\\\frac{1}{G_{vh}}
\end{array}
\right\} 
= \left\{\begin{array}{c}
\varepsilon_{z}/\sigma_{z}\\\varepsilon_{y}/\sigma_{z}\\\varepsilon_{y}/\sigma_{y}\\\varepsilon_{x}/\sigma_{y}\\\varepsilon_{z1'}/\sigma_{z1'}\\\varepsilon_{z2'}/\sigma_{z2'}\\\varepsilon_{z3'}/\sigma_{z3'}\end{array}\right\}
\end{equation}
In Eq. \ref{Eq: over}, subscripts $1$, $2$, and $3$ are used to indicate the three specimens with $\theta= 30^\circ, 45^\circ$, and $60^\circ$. The best-fit of the five independent elastic constants can be determined by applying the least-squares method to Eq. \ref{Eq: over}. 

\subsection{Seismic velocities and elastic stiffness matrix}\label{Sec: 2P3}
The seismic velocities of the Marcellus shale samples were measured in this work, and can be related to the components of the dynamic stiffness matrix. For a general elastic medium, seismic waves are divided into compressional waves (P-waves) and shear waves (S-waves). When a shear wave enters a transversely isotropic medium, it splits into two orthogonal components: one wave vibrating parallel to the plane of isotropy (SH-wave) and another wave vibrating in a plane orthogonal to that plane (SV-wave). In total, three seismic velocities associated with the P-wave, SH-wave, and SV-wave, respectively, were measured, and these can be expressed as functions of the components of the elastic stiffness matrix \cite{mavko2009rock,thomsen1986weak,byun1986apparent}: 

\begin{subequations}\label{Eq:vp}
	\begin{align}
	2\rho V_P^2(\theta)& = C_{11}\sin^2\theta + C_{33}\cos^2\theta + C_{44} + \sqrt{M(\theta)} \\
	2\rho V_{SV}^2(\theta)& = C_{11}\sin^2\theta + C_{33}\cos^2\theta + C_{44} - \sqrt{M}(\theta) \\
	\rho V_{SH}^2(\theta)& = C_{66}\sin^2\theta + C_{44}\cos^2
	\theta
	\end{align}
\end{subequations}
where
\begin{equation}
\resizebox{.9\hsize}{!}{$M(\theta) = \left[\left(C_{11}-C_{44}\right) \sin^2(\theta) -  \left(C_{33}-C_{44}\right)\cos^2(\theta)\right]^2 
+ \left(C_{13} + C_{44}\right)^2 \sin^2(2\theta)$}
\end{equation}

In the above equation, $\rho$ is the density of material; $V_P$, $V_{SV}$ and $V_{SH}$ are the velocities associated with the P-wave, SV-wave, and SH-wave, respectively; $\theta$ is anisotropy angle; $C_{ij}$ are the components of the elastic stiffness matrix. 

In this work, P-wave velocities of specimens with various angles of bedding plane inclination, as well as SV-wave and SH-wave velocities of the specimens with $\theta = 0^\circ$ and $90^\circ$, were measured. Similar to Section \ref{Sec: EC}, the five components of the stiffness matrix can be determined by applying the least square method according to Eq. \ref{Eq:vp}.

\section{Experimental setup}	
\subsection{Sample preparation}

\begin{figure}
	\begin{center}
		\includegraphics[width = 0.9\textwidth]{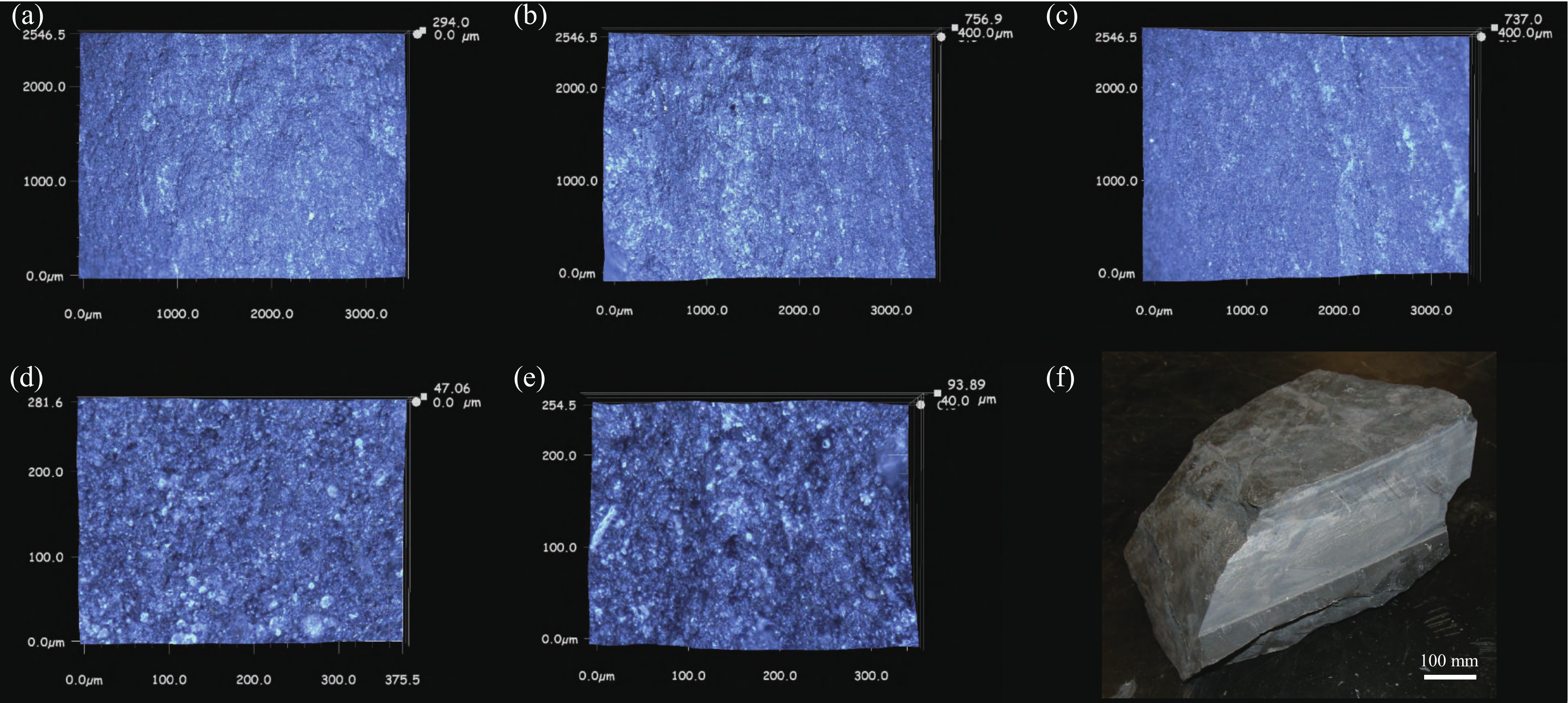}\\
		\caption{(a)-(e) Micrographs of several representative samples oriented with the bedding direction from top to bottom. (f) Shale block from Marcellus outcrop. }
%
%
		\label{Fig: MAT}
	\end{center}
\end{figure}

The shale materials used in this work were taken from the outcrops of the Marcellus Formation. The blocks are black and compact, and featured by alternating light and dark layers representing fine lamination or bedding formed by the accumulation of sediments (see Fig. \ref{Fig: MAT}f). The material also shows clear evidence of micro-scale heterogeneity revealed by an internal structure appearing under a microscope, as shown in Fig. \ref{Fig: MAT}a-\ref{Fig: MAT}e. Visual inspection showed that the materials are free of surface cracks and voids. The sample is considered to be dry as water content by mass measured by following ASTM D2216 is less than 0.2\%. The average mass density is $2558~\text{kg}/\text{m}^3$. 

\begin{figure}
	\begin{center}
		\includegraphics[scale=0.36, trim=0mm 0mm 0mm 0mm, clip ]{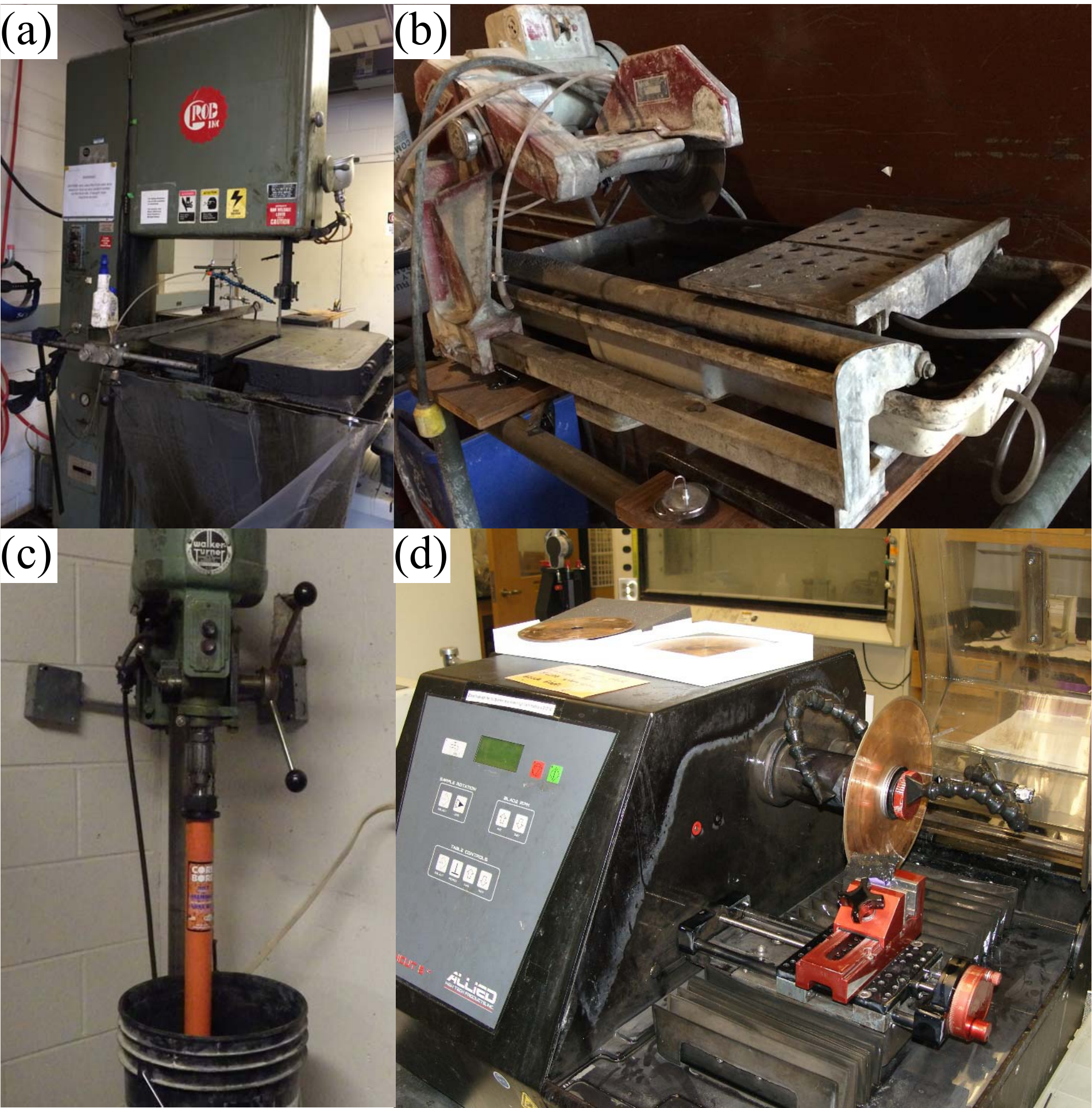}\\
		\caption{Sample preparation tools: (a) Band saw for cutting cube sample; (b) Circular saw for cutting large blocks into small pieces; (c) Directional coring for disc sample; (d) Multi-purpose precision sectioning saw for samples in direct tension tests.}
		\label{Fig:Tools}
	\end{center}
\end{figure}

The large Marcellus shale block was first cut into small chunks by using a table tile saw with a diamond-coated blade as shown in Fig. \ref{Fig:Tools}b. The small specimens were cut from the chunks into various shapes for different tests. 


For seismic velocity measurement, cubic specimens with a nominal dimension of $25$ mm $\times$ $25$ mm $\times$ $25$ mm were prepared. A band saw (Fig.~\ref{Fig:Tools}a) was used to prepare the cubic specimens with anisotropy angle $\theta$ varying between $0^{\circ}$ (bedding plane perpendicular to the loading direction) and $90^{\circ}$ (bedding parallel to loading direction). Specimens with seven different anisotropy angles of $0^{\circ}$, $15^{\circ}$, $25^{\circ}$, $45^{\circ}$, $65^{\circ}$, $75^{\circ}$ and $90^{\circ}$ were prepared.

\begin{figure}[h]
	\begin{center}
		\includegraphics[width = 0.55\textwidth]{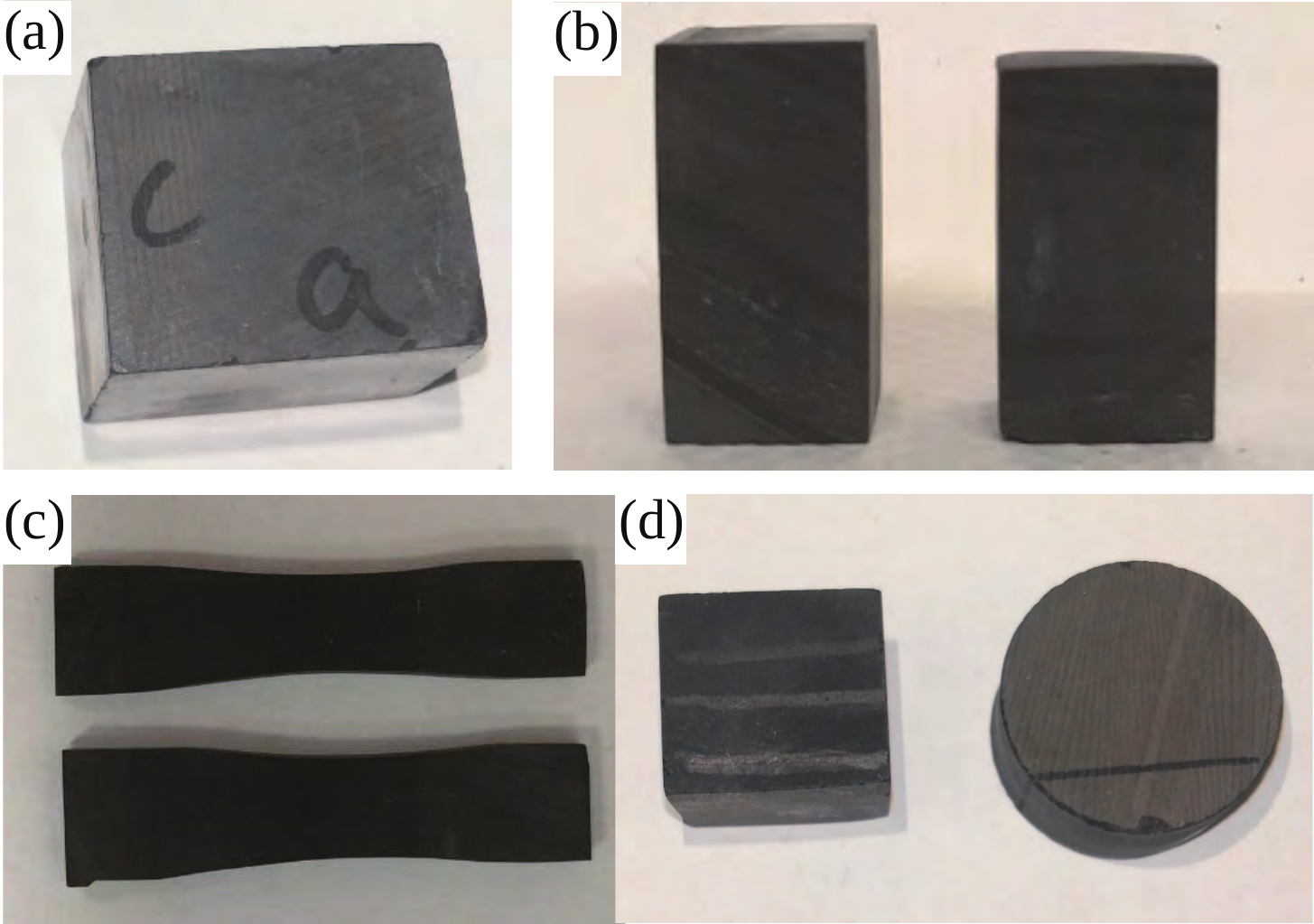}
		\caption{Typical specimens for (a) seismic velocity measurement; (b) uniaxial compression test; (c) direct tension tests; (d) Brazilian test.}
		\label{Fig:Specimens}
	\end{center}
\end{figure}

For uniaxial compression tests, the cuboid specimens with a typical dimension of $12$ mm (length) $\times$ $12$ mm (width) $\times$ $20$ mm (height) were prepared by using the band saw and a TechCut 5$^{\text{TM}}$ precision sectioning machine (Fig. \ref{Fig:Tools}d). The specimens were classified into five groups according to their anisotropy angles of $0^{\circ}$, $30^{\circ}$, $45^{\circ}$, $60^{\circ}$, and $90^{\circ}$. The typical specimens are shown in Fig.~\ref{Fig:Specimens}b.

For direct tension tests, the small coupons of 2 mm thickness were prepared by using the precision sectioning machine. The in-plane dimension is $40$ mm (length) $\times$ $8$ mm (width). To assist gripping under uniaxial loading, the specimens were prepared in such a way that the gage region was necked to generate a so-called waisted or dog-bone profile. Similar to other types of specimens, the samples for the direct tension tests were also categorized according to five different anisotropy angles of $0^{\circ}$, $30^{\circ}$, $45^{\circ}$, $60^{\circ}$ and $90^{\circ}$.

For the Brazilian tensile tests, disc-shaped specimens, each with a nominal diameter, $d$, of $38$ mm and a nominal height, $h$, of $19$ mm, were prepared by using a laboratory directional coring system (Fig. \ref{Fig:Tools}c). The shale block was clamped to prevent any unwanted movement during the coring process. The speed of coring was constant to avoid any irregularities or defects on the cutting surface. The specimens were cored parallel to the bedding plane. 
                      
\begin{figure}                                                
	\begin{center}
		\includegraphics[width = 0.6\textwidth]{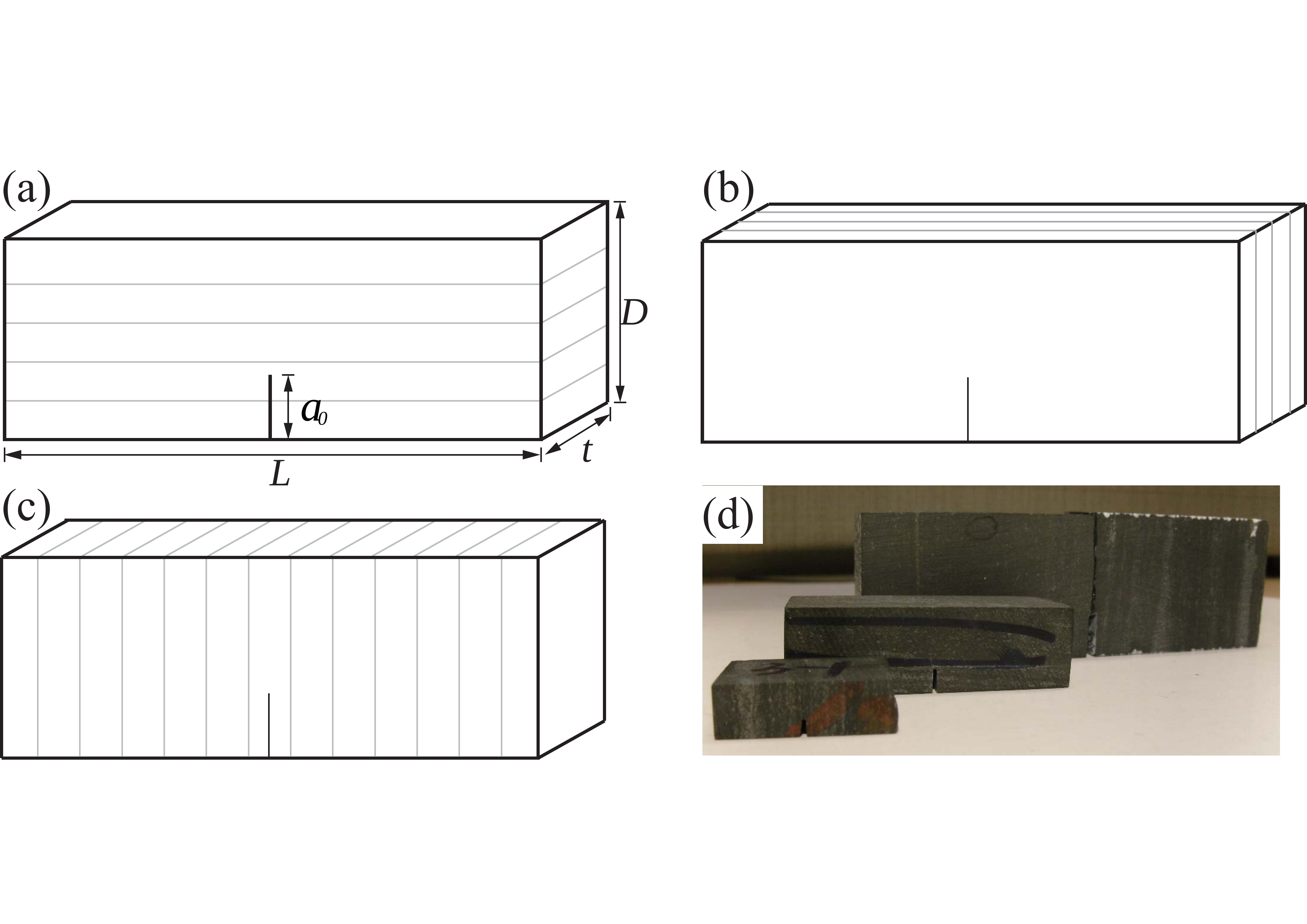}
		\caption{Sketch of the specimens with three principal notch orientations: (a) arrester, (b) divider, and (c) short-transverse. (d) Typical specimens with increasing size. }
		\label{Fig: SESamples}
	\end{center}
\end{figure}

For the fracture tests, Three-Point-Bending (TPB) specimens with length $L$, depth $D$, thickness $t$, and notches of length $a_0$ were prepared by using the precision sectioning machine (Fig. \ref{Fig:Tools}d). A diamond wafering blade with thickness of 0.36 mm was used to machine the notches. Following the pioneering work by Schmidt \cite{schmidt1977fracture} and Chong et al. \cite{chong1987fracture}, the specimens were made in such a way that the notches were aligned with one of three principal orientations with respect to the plane of isotropy, referred to as \textit{arrester},\textit{divider}, and \textit{short-transverse}, as depicted in Fig. \ref{Fig: SESamples}a, \ref{Fig: SESamples}b, and \ref{Fig: SESamples}c, respectively. In order to explore the size effect, specimens of scaled planar dimensions and constant thickness were prepared for each specimen configuration. Three sizes with geometric ratio of 4:2:1, namely large, medium, and small, were considered. The depths of the three types of specimens were around 25 mm, 12.5, and 6.25 mm, respectively, while the thicknesses were around 14 mm. The typical TPB specimens with varying sizes are shown in Fig. \ref{Fig: SESamples}d. 

\subsection{Seismic velocity measurements}
\begin{figure}                                                 
	\begin{center}
		\includegraphics[width = 0.55\textwidth]{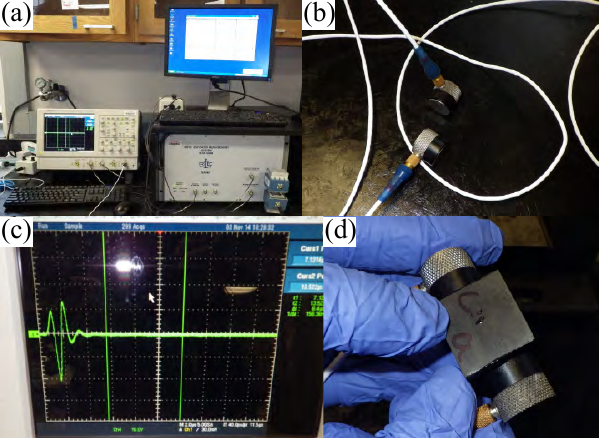}
		\caption{The measurement system used to determine the seismicity of each sample. (a) RAM-5000 computer controlled ultrasonic system; (b) transmitter and receiver; (c) oscilloscope; and (d) experimental sample. }
		\label{2.6}
	\end{center}
\end{figure}

A seismic velocity measurement system was used to determine the compressive wave (P-wave) and shear wave (SH-wave and SV-wave) velocities of the Marcellus shale samples, as shown in Fig.~\ref{2.6}. The system consists of several components, including the pulser/receiver, transducer, and a display device. A pulser/receiver is an electronic device that can produce high voltage electrical pulses. Driven by the pulser, the transducer generates high frequency ultrasonic energy. The sound energy is introduced and propagates through the materials in the form of stress waves. The wave signal is transformed into an electrical signal by the transducer and is displayed on a screen. The velocity of the wave is calculated as the distance that the signal traveled divided by the travel time. Longitudinal and shear wave transducers were used for P-wave and S-wave velocity measurements respectively, and the polarization direction of the shear wave transducers was aligned perpendicular to the bedding plane of the material separately in order to induce SH-waves and SV-waves. 

\subsection{Uniaxial compression tests}
\begin{figure}
	\begin{center}
		\includegraphics[scale=0.56, trim=72mm 70mm 0mm 30mm, clip]{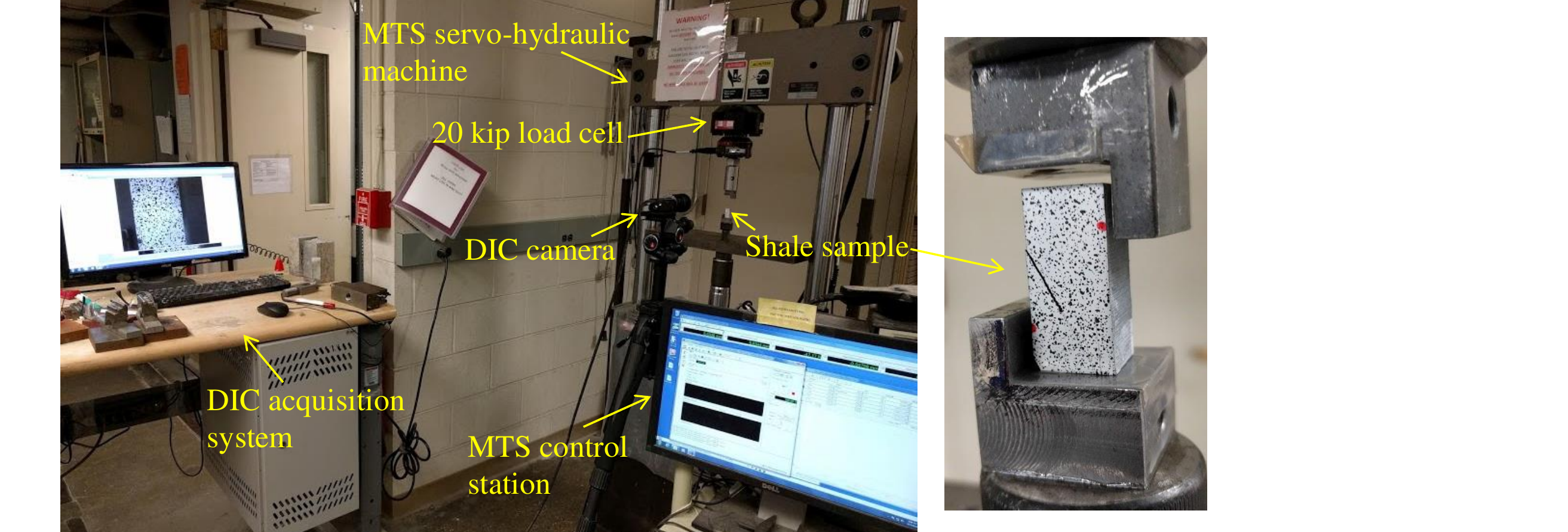}\\
		\caption{The experimental setup for uniaxial compression tests.}
		\label{2.7}
	\end{center}
\end{figure}

A MTS material testing system was used to complete the uniaxial compression tests. The system consists of a servohydraulic load frame, a hydraulic pump unit, a controller, a load cell operating in the 20 kip (88.96 kN) range, and a computer, as shown in Fig.~\ref{2.7}. The prepared cuboid specimens were placed between two loading platens. A Teflon sheet was stuck to the platens to reduce the friction between the specimens and the steel platens (friction coefficient $\approx 0.04$). All specimens were loaded in compression up to failure at a constant displacement rate of $0.0025$ mm/s. In the loading process, the force was recorded. The nominal compressive stress was calculated as $\sigma=F/A$, where $F$ is the force applied on the specimen; $A$ is the initial area of the specimen cross section. The UCS of the material was determined by loading specimens to failure and calculating the UCS as the maximum nominal compressive stress.

During the test, strain was measured by using the Digital Image Correlation (DIC) technique. DIC is an optical method which uses a mathematical correlation between digital images taken while samples are tested to deduce the displacement field \cite{solutions2009vic}. In this work, the Vic-2D system, consisting of a computer software and a digital camera (shown in Fig.~\ref{2.7}), was used to provide two-dimensional strain maps of an entire planar specimen surface. 

The axial and lateral strains were averaged over the gage region (8 $\times$ 10 mm). In order to synchronize the data of stress and strain, the MTS machine and DIC system were started at the same time. Consequently, the stress and strain data obtained simultaneously could be correlated and used to investigate the elastic behaviors of the material.

\subsection{Direct tension tests}
\begin{figure}
	\begin{center}
		\includegraphics[scale=0.5, trim=20mm 65mm 0mm 25mm, clip]{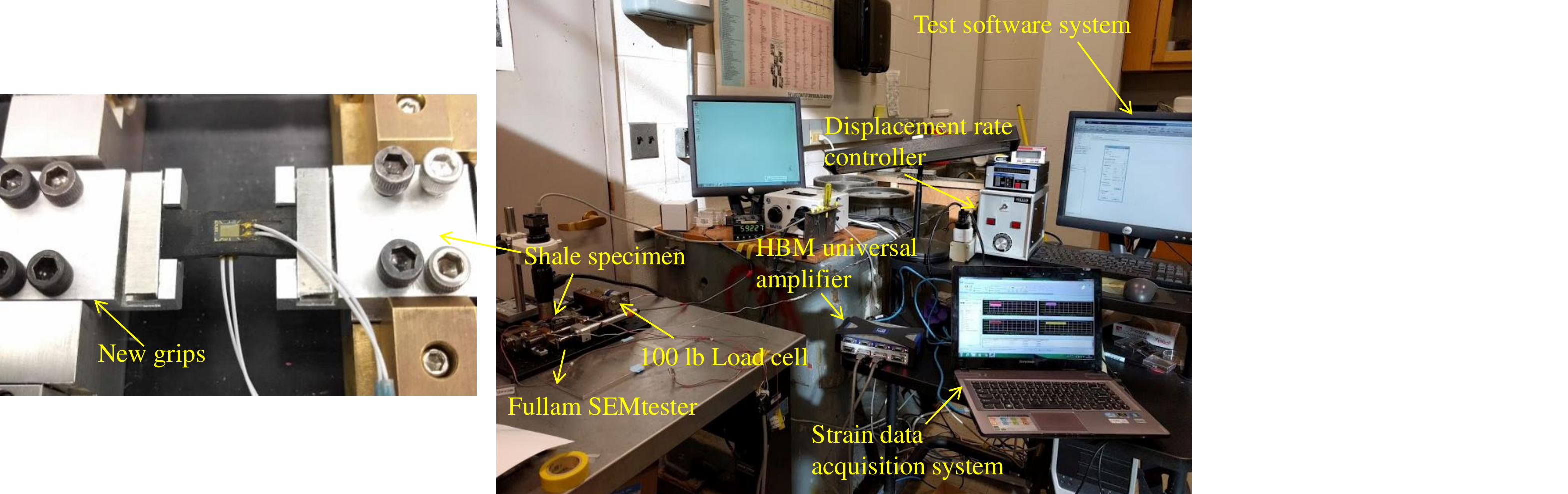}\\
		\caption{ Experimental setup for direct tension tests}
		\label{Fig:TensileSetup}
	\end{center}
\end{figure}

The prepared waisted specimens were loaded by using a Fullum SEMtester with a load cell operated in a 100 lb (444.82 N) range under displacement control.  A constant displacement rate of $0.000018$ mm/s was applied. To ensure the uniaxial loading condition, special grips with a T-shaped groove (see Fig.~\ref{Fig:TensileSetup}) were designed and used in the tests to minimize the unwanted effect of bending or torsion introduced during test operations. Two aluminum sheets were stuck to the upper and bottom of the specimens on each side, and the flanges were fitted into the grooves of the grips to introduce the gripping force. During the test, the specimens were stretched up to failure, and the data of load and displacement were recorded. The nominal tensile stress is again defined as $\sigma=F/A$, where $F$ is again the uniaxial force applied to the specimen; $A$ for the tensile tests is the initial area of the specimen cross section calculated at the waisted region. By loading to failure, the Direct Tensile Strength (DTS) of the material was determined as the maximum nominal tensile stress.

Strain gages were used to measure the strain during direct tension tests. Two strain gages with 3 mm active gage length for axial strain measurement and 1.5 mm for lateral strain measurement were mounted on the center of each specimen surface. The strain gages were connected to a HBM universal amplifier, and the strain data were recorded during the test. The entire experiment setup is shown in Fig~\ref{Fig:TensileSetup}.

\subsection{Brazilian tests}
The Brazilian tests were conducted with a $1000$ kip MTS testing system using displacement control. To investigate the directionally dependent response, the specimens were rotated such that the bedding plane has a angle with respect to the loading direction, i.e. anisotropy angle, as illustrated in Fig. \ref{Fig:Anisotropy}b.  The specimens were loaded up to failure at a constant displacement rate of $0.003$ mm/s. After the failure of specimens, pieces of the failed rock specimens were collected to investigate the relationship between the bedding plane and the failure plane. 

The expression of indirect tensile strength from Brazilian test for anisotropic rocks was derived analytically by Claesson and Bohloli \cite{claesson2002brazilian}, and can be written as
\begin{equation}\label{Eq:BTS}
\resizebox{0.9\hsize}{!}{$\text{BTS} = \frac{2P}{\pi dh}\left[ \left( \sqrt[4]{E_h/E_v}\right)^{\cos(2\theta)} - \frac{\cos(4\theta)}{4}(b-1) \right], \quad
b = \frac{\sqrt{E_h E_v}}{2}\left( \frac{1}{G_{vh}} - \frac{2 \nu_{vh}}{E_v} \right)$}
\end{equation}
where $P$ is the load at failure; $d$ is the diameter of the test specimen; and $h$ is the thickness.

\subsection{Size effect test on three-point-bending specimens}
The prepared TPB specimens were placed on two supporting pins with the support span, being 74, 37, and 18.5 mm for large, medium, and small size, respectively, and were loaded vertically under symmetrical three-point bending. The tests were conducted using displacement control on a Mini-Tester with  closed-loop control and a load cell operating in the 200 lb (889.64 N) range. Constant displacement rate of 0.1, 0.05, and 0.025 mm/min were used for large, medium, and small specimens, respectively, to ensure the same strain rate for all investigated specimens. The load-point displacements and loads were recorded during the tests. In total, 27 tests were conducted with three tests for each specimen size and configuration. 

\section{Results and analysis}	
\subsection{Results of seismic velocity measurement}\label{Sec: 4P1}
\begin{figure}
	\begin{center}
		\includegraphics[scale=0.7]{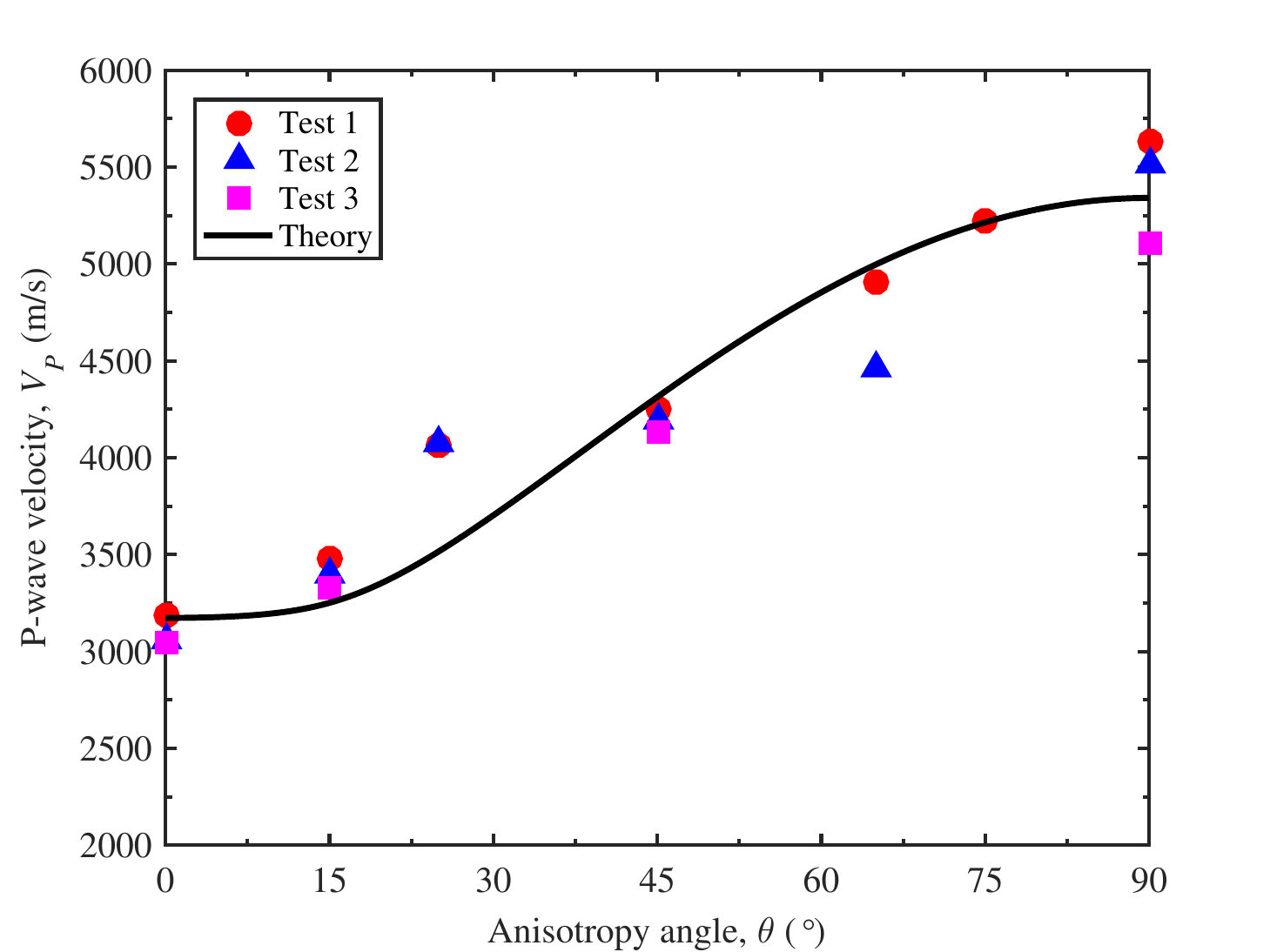}
		\caption{P-wave velocity variation with respect to anisotropy angle}
		\label{Fig:vp}
		\end{center}
\end{figure}

The calculated P-wave velocity for each specimen is plotted against the specimen anisotropy angle in Fig.~\ref{Fig:vp}. It can be seen that the P-wave velocity of the material increased with the anisotropy angle, and varied from $3104$ m/s to $5481$ m/s. The maximum values occurred when the direction of the longitudinal wave propagation was parallel to the plane of isotropy, or bedding plane, which was obtained with the specimen with $\theta=90^{\circ}$. The minimum value occurred at the anisotropy angle of $0^{\circ}$, where the direction of the longitudinal wave propagation was perpendicular to the plane of isotropy. The anisotropy ratio of the P-wave velocity, defined as $V^{max}_P /V^{min}_P$, is about $1.77$. 

SH-wave and SV-wave velocities were collected for the specimens with anisotropy angles of $0^\circ$ and $90^\circ$,  and were determined by averaging three independent measurements. For the specimens with  $\theta = 90^\circ$, the average SH-wave velocity was calculated as $2980$ m/s, and the average SV-wave velocity was $2855$ m/s. For the specimens with  $\theta = 0^\circ$, the shear wave propagates along the symmetric axis, and the SH-wave and SV-wave become identical according to Eq. \ref{Eq:vp}b and \ref{Eq:vp}c. The shear wave velocity of specimen with $\theta = 90^{\circ}$ averages at $2819$ m/s. 
%
%

The dynamic elastic constants were derived from the velocity measurements and the density value of the material according to Eq. \ref{Eq:vp}a by using the lease-square method. The results are reported in Table~\ref{stiffness}. The obtained constants are also used to obtain a prediction of $V_P$, as represented by the black solid line in Fig. \ref{Fig:vp}. The accuracy of the prediction was quantified by the Mean Absolute Percentage Error (MAPE), which is defined by the formula:
\begin{equation}
\text{MAPE} = \frac{100}{n}\sum_{t=1}^n \left| \frac{E_t - F_t}{E_t} \right|
\end{equation}
where $E_t$ is the experimental value; $F_t$ is the predicted one. The calculated MAPE value for P-wave velocity is 1.5\%, and the agreement between the experiments and predictions is generally acceptable as one can observe from Fig. \ref{Fig:vp}.

In addition, the apparent dynamic Young's moduli can be calculated from the stiffness matrix whose components are reported in Table \ref{stiffness}. They are $E_v^{d} = 22.81$ GPa and $E_h^{d} = 53.95$ GPa for dynamic modulus in vertical and horizontal direction, respectively. Compared to the static Young's moduli measured from either uniaxial compression or direct tension tests, which will be presented in the following sections, the dynamic Young's moduli calculated from the seismic velocity measurements are around 50\% higher. The difference observed in this work is in agreement with previous publications on static-dynamic relations for rocks \cite{ide1936comparison,eissa1988relation,sone2013mechanical}. 
%
%


\begin{table}
	\caption{Components of dynamic elastic stiffness matrix calculated from seismic velocity measurements.}
	\label{stiffness}
	\renewcommand{\arraystretch}{1.0}
	\begin{center}
		\resizebox{0.6\textwidth}{!}{%
        \begin{tabular}{ccccc}
			\hline
			$C_{11}$(GPa)&$C_{33}$(GPa)&$C_{13}$(GPa)&$C_{44}$(GPa)&$C_{66}$(GPa)\\
			\hline
			72.98&25.76&-12.17&20.72&22.74\\
			\hline			
		\end{tabular}}%
	\end{center}
\end{table}

\subsection{Results of uniaxial compression test}
\subsubsection{Five elastic constants}
Given the assumption of transverse isotropy, one can derive the five static elastic constants for Marcellus shale from the uniaxial compression tests, as reported in the first row of Table~\ref{tab: Constants}. These results were calculated from the stress and strain measurements on all investigated specimens, with anisotropy angles $\theta= 0^{\circ}, 30^{\circ}, 45^{\circ}, 60^{\circ}$, and $90^{\circ}$, and by using the least-squares method according to Section \ref{Sec: EC}. In the least-squares equations, the values of $\varepsilon/\sigma$ were calculated according to the secant elastic modulus method at $50\%$ of maximum strain. 

The table also reports the shear modulus in the plane normal to the plane of isotropy, $G_{SV}$,calculated by using Saint-Venant's empirical equations \cite{saint1863distribution,cho2012deformation}. This empirical approximation can be written as
%
%
%
%
\begin{equation}
	\frac{1}{G_{SV}} = \frac{1}{E_h} + \frac{1+2\nu_{vh}}{E_v}
\end{equation}
The relative difference of the shear modulus determined by these two methods is also calculated and reported in Table~\ref{tab: Constants}.

\begin{table}[htbp]
	\centering
	\caption{Five elastic constants for Marcellus shale based on uniaxial compression tests.}
    \resizebox{1.0\textwidth}{!}{%
	\begin{tabular}{lccccccc}
		\hline
		& $E_v$ (GPa) & $E_h$ (GPa)& $\nu_{vh}$ & $\nu_{hh}$ & $G_{vh}$ (GPa) & $G_{SV}$ (GPa) & $|G_{vh}-G_{SV}|/G_{vh}$\\
		\hline
		Compression & 16.12 & 37.72 & 0.35  & 0.25  & 6.87 & 7.58 & 0.103\\
		Tension & 11.50  & 37.06 & 0.33  & 0.18  & 6.40 & 5.84 & 0.088 \\
		\hline
	\end{tabular}}%
	\label{tab: Constants}%
\end{table}%

\subsubsection{Compressive modulus}
The apparent Young's moduli of the specimens under compression, denoted by $E_c$, were measured from the uniaxial compression tests, and these are reported in Table \ref{tab:compression} from column two to six. The results are categorized according to specimen anisotropy angle, and in each case, the mean value and Standard Deviation (SD) are calculated. It can be seen that the reported moduli varied from $14.50$ to $45.80$ GPa and were strongly related to anisotropy angle. The maximum compressive modulus occurred at $\theta=90^{\circ}$, while the minimum occurred at $\theta=0^{\circ}$. The anisotropy ratio of the compressive modulus, defined by $\bar{E}_c^{max} / \bar{E}_c^{min}$, is about $2.24$. 

\begin{figure}[h]
	\begin{center}
		\includegraphics[scale = 0.52,trim=20mm 0mm 0mm 0mm, clip]{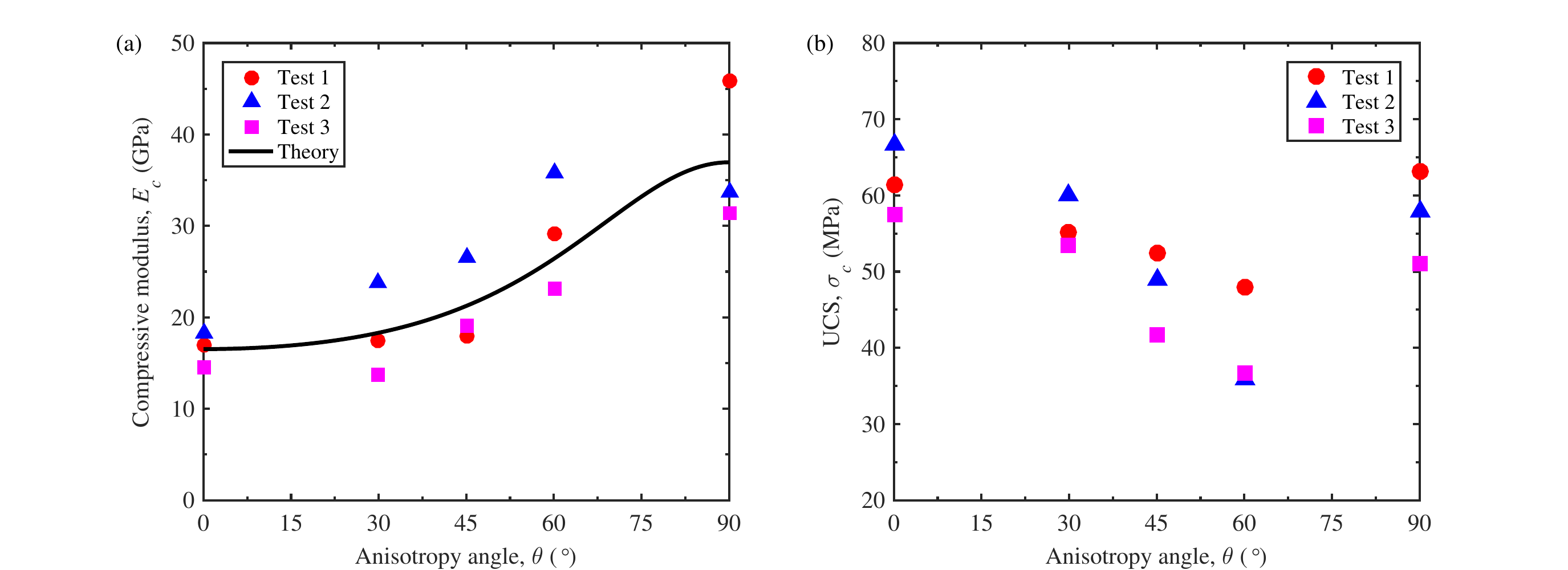}
		\caption{Variation of (a) compressive modulus (b) Uniaxial Compressive Strength (UCS) with anisotropy angle.}
		\label{3.4}
	\end{center}
\end{figure}

Fig. \ref{3.4}a shows the variation of the measured compressive modulus with anisotropy angle. One can observe that $E_c$ increased with an increase of $\theta$. The solid line represents the predicted Young's moduli according to Eq. \ref{Eq: E33}, based on the assumption of transverse isotropy, and the elastic constants under compression reported in Table \ref{tab: Constants}. It can be seen that the measured and predicted values of $E_c$ are close, and the resultant MAPE value is 3.1\%. 

\begin{table}[htbp]
	\centering
	\caption{Results of apparent elastic moduli and Uniaxial Compressive Strength (UCS) from uniaxial compression tests}
    \resizebox{0.9\textwidth}{!}{%
	\begin{tabular}{lcccccccccc}
		\hline
          $\theta $& \multicolumn{3}{l}{$E_c$ (GPa)} & \multicolumn{1}{l}{Mean} & \multicolumn{1}{l}{SD} & \multicolumn{3}{l}{UCS (MPa)} & \multicolumn{1}{l}{Mean} & \multicolumn{1}{l}{SD} \\
          \hline
          0     & 16.90 & 18.19 & 14.50 & 16.53 & 1.53  & 61.42 & 66.58 & 57.47 & 61.82 & 3.73 \\
          30    & 17.50 & 23.80 & 13.67 & 18.32 & 4.18  & 55.17 & 60.00 & 53.43 & 56.20 & 2.78 \\
          45    & 17.91 & 26.62 & 19.00 & 21.18 & 3.87  & 52.38 & 48.91 & 41.74 & 47.68 & 4.43 \\
          60    & 29.12 & 35.80 & 23.12 & 29.35 & 5.18  & 48.00 & 35.80 & 36.70 & 40.17 & 5.55 \\
          90    & 45.80 & 33.73 & 31.36 & 36.96 & 6.32  & 63.20 & 57.88 & 51.10 & 57.39 & 4.95 \\
		\hline
	\end{tabular}}%
	\label{tab:compression}%
\end{table}%

\subsubsection{Compressive strength}
The calculated UCS of the specimens are reported in Table~\ref{tab:compression} from columns seven to eleven, and also plotted in Fig. \ref{3.4}b. The measured UCS values vary from $41.00$ to $66.58$ MPa. With an increase of anisotropy angle, the UCS decreases at first, followed by an increase after $\theta \geq 60^{\circ}$. The material exhibited the maximum UCS either at $\theta=0^{\circ}$ or $90^{\circ}$, and the minimum value occurred at $\theta = 60^{\circ}$. The anisotropy ratio of uniaxial compressive strength, defined by UCS$^{max}/$UCS$^{min}$, is about $1.54$.  This anisotropy ratio is lower compared to the anisotropy ratio of the compressive modulus. This disparity, and the stronger anisotropy with respect to deformability, might result from the organized distribution of minerals and compliant organic materials.

\begin{figure}
	\begin{center}
		\includegraphics[width = 0.8\textwidth]{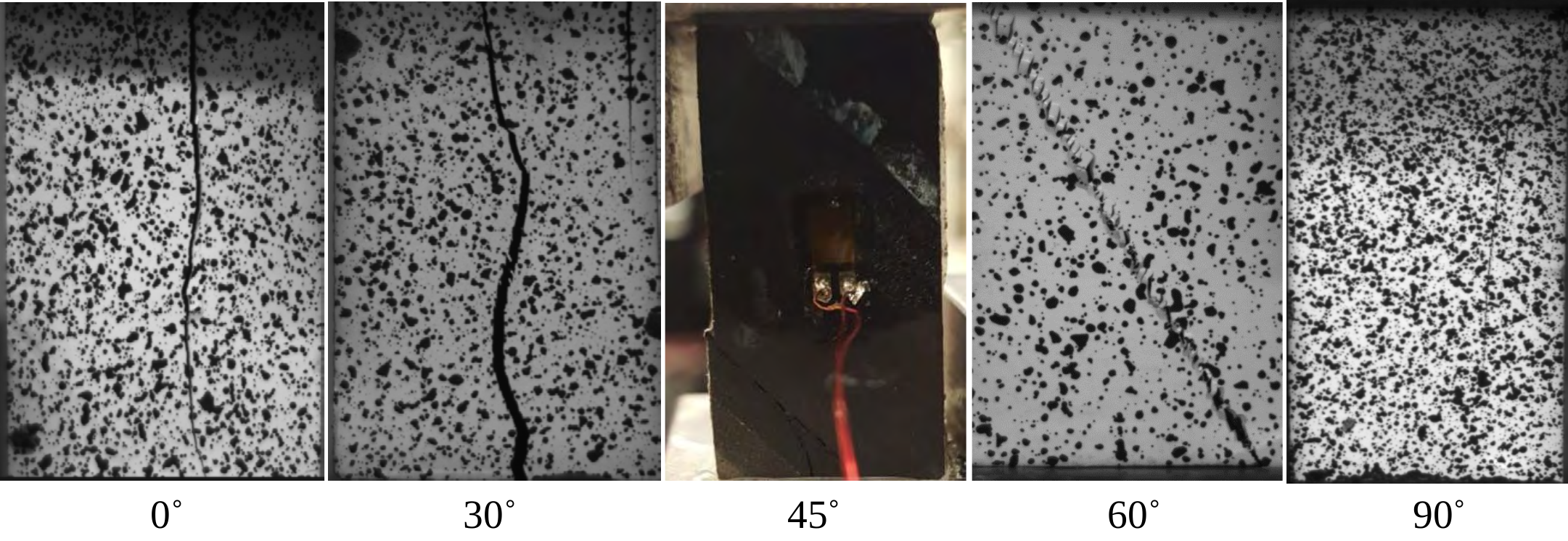}
		\caption{Photograph of specimens after failure in uniaxial compression tests.}
		\label{Fig: UCFailure}
	\end{center}
\end{figure}

Fig.~\ref{Fig: UCFailure} shows photographs of failed specimens in the uniaxial compression tests, highlighting typical modes of failure as they varied with the anisotropy angle. A straight failure path was observed for the specimens with $\theta=0^{\circ}$ and $\theta=90^{\circ}$, indicating axial splitting as the type of failure. Although similar failure modes were observed, the failure plane coincided with the bedding plane for the specimen with $\theta=90^{\circ}$, while it was perpendicular to bedding with $\theta=0^{\circ}$. For the specimens with $\theta=45^{\circ}$ and $60^{\circ}$, the failure path was mostly aligned with the bedding direction, indicating a shear failure mode. For the specimen with $\theta = 30^{\circ}$, the failure path was tortuous, and observed to be both along and crossing the bedding plane, indicating a mixed splitting and shear failure mode. 

According to the ``plane of weakness'' model \cite{jaeger1960shear, mclamore1967mechanical} and subsequent numerical studies \cite{li2017multiscale},  the UCS tends to be lower when the shear failure plane corresponds to the plane of weakness, which is the case for $\theta = 45^\circ$ and $60^\circ$ as shown in  Table~\ref{tab:compression}  and Fig. \ref{3.4}b. Given that the failure path was aligned with the bedding direction for the specimens with $\theta=45^{\circ}$ and $60^{\circ}$, the widely accepted assumption of bedding planes being planes of weakness is believed to be valid for the sample of Marcellus shale studied. 

\subsection{Results of direct tension tests}
\subsubsection{Five elastic constants}
Similar to the analysis of the experimental data from the uniaxial compression tests, the five elastic constants for the material under direct tension were determined, and the results are reported in the second row of Table \ref{tab: Constants}. One may note that the values that the moduli inferred from tension and compression tests different somewhat, a difference that is discussed in Section \ref{discuss_ten_comp}. Also, one can observe that the shear modulus normal to the plane of isotropy calculated by using the Saint-Venant's empirical equations is close to the one determined by using the least-squares method for both the uniaxial compression and direct tension tests, with the relative difference around 0.1. The mutual validation of these two methods indicates that the theory of elasticity for transversely isotropic materials is applicable. 
%
%
%
%
\subsubsection{Tensile modulus}
\begin{figure}
	\begin{center}
		\includegraphics[scale = 0.52, trim=20mm 0mm 0mm 0mm, clip]{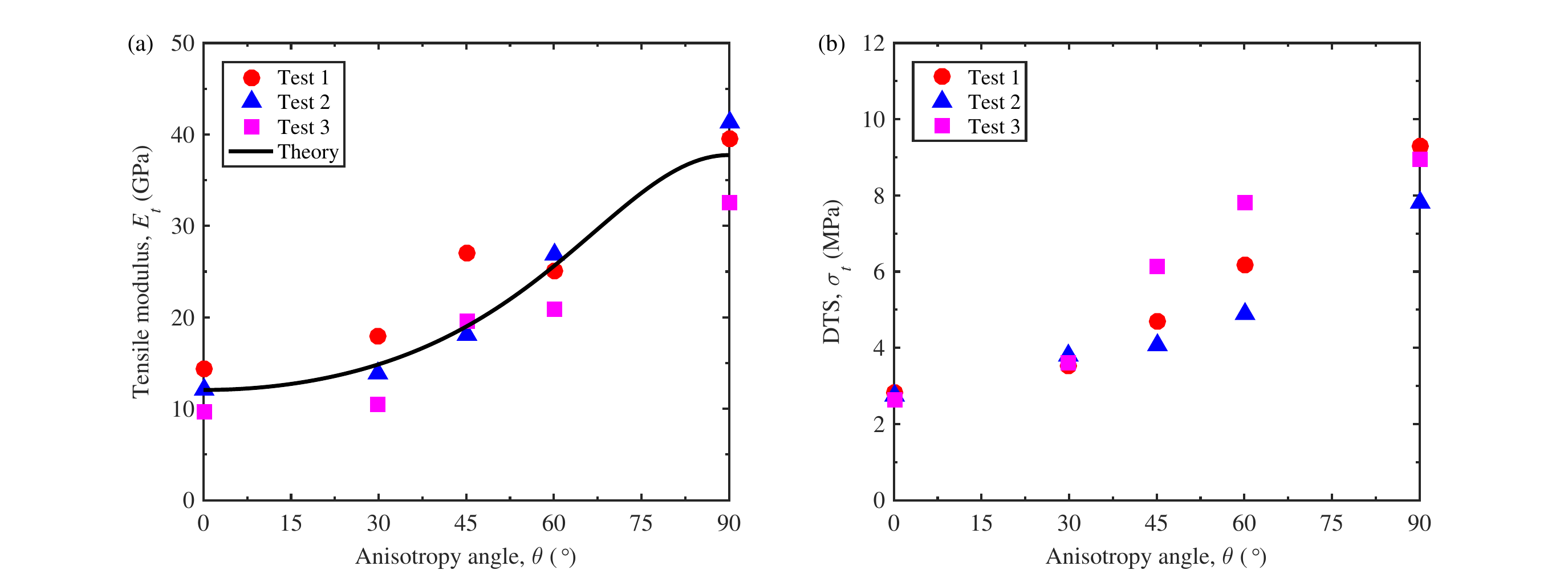}
		\caption{Variation of (a) tensile modulus (b) and Direct Tensile Strength (DTS) with anisotropy angle.}
		\label{Fig: Tension}
	\end{center}
\end{figure}

Columns two to six in Table \ref{tab:Tension} give the apparent elastic moduli of the specimens measured in the direct tension tests, denoted by $E_t$. The variation of $E_t$ with the specimen anisotropy angle is also plotted in Fig. \ref{Fig: Tension}a. It can be observed that the measured tensile modulus varied from $9.90$ to $41.25$ GPa and increased with anisotropy angle. The maximum tensile modulus occurred at  $\theta=90^{\circ}$, while the minimum one occurred at  $\theta=0^{\circ}$. The anisotropy ratio of the tensile modulus, defined by $\bar{E}_t^{max} / \bar{E}_t^{min}$, is about $3.13$, which is slightly larger than that of the compressive modulus. 

The apparent Young's moduli predicted by the theory of elasticity under the assumption of transverse isotropy, represented by the solid line in Fig. \ref{3.4}a, were also compared with the experimental data. Good agreement between the predicted and measured $E_t$ can be observed, and the MAPE value is reported to be 1.9\%.

\begin{table}[htbp]
	\centering
	\caption{Results of apparent elastic moduli and Direct Tensile Strength (DTS) from direct tension tests}
	\resizebox{0.9\textwidth}{!}{%
    \begin{tabular}{lcccccccccc}
		\hline
		$\theta$& \multicolumn{3}{l}{$E_t$ (GPa)} & \multicolumn{1}{l}{Mean} & \multicolumn{1}{l}{SD} & \multicolumn{3}{l}{DTS (MPa)} & \multicolumn{1}{l}{Mean} & \multicolumn{1}{l}{SD} \\
		\hline
		0     & 14.44 & 12.15 & 9.60  & 12.06 & 1.98  & 2.84  & 2.73  & 2.63  & 2.73  & 0.09 \\
		30    & 17.92 & 13.86 & 10.40 & 14.06 & 3.07  & 3.53  & 3.79  & 3.62  & 3.65  & 0.11 \\
		45    & 27.04 & 18.10 & 19.59 & 21.58 & 3.91  & 4.71  & 4.09  & 6.12  & 4.97  & 0.85 \\
		60    & 25.09 & 26.92 & 20.90 & 24.30 & 2.52  & 6.16  & 4.9   & 7.8   & 6.29  & 1.19 \\
		90    & 39.45 & 41.25 & 32.54 & 37.75 & 3.75  & 9.29  & 7.82  & 8.96  & 8.69  & 0.63 \\
		\hline
	\end{tabular}}%
	\label{tab:Tension}%
\end{table}%

\subsubsection{Tensile strength}

The measured DTS of the specimens are reported in Table \ref{tab:Tension} from column seven to eleven. The strength values are also plotted against the anisotropy angle in Fig. \ref{Fig: Tension}b. The measured DTS  varies from $2.63$ to $9.28$ MPa, with the maximum value occurring at $\theta=90^{\circ}$ and the minimum one at $\theta=0^{\circ}$. It can be noted that the tensile strength gradually increased with an increase of anisotropy angle. The anisotropy ratio of DTS, defined by DTS$^{max}$/DTS$^{min}$, is $3.18$. 

\begin{figure}
	\begin{center}
		\includegraphics[scale = 0.5,  trim=50mm 72mm 0mm 35mm, clip]{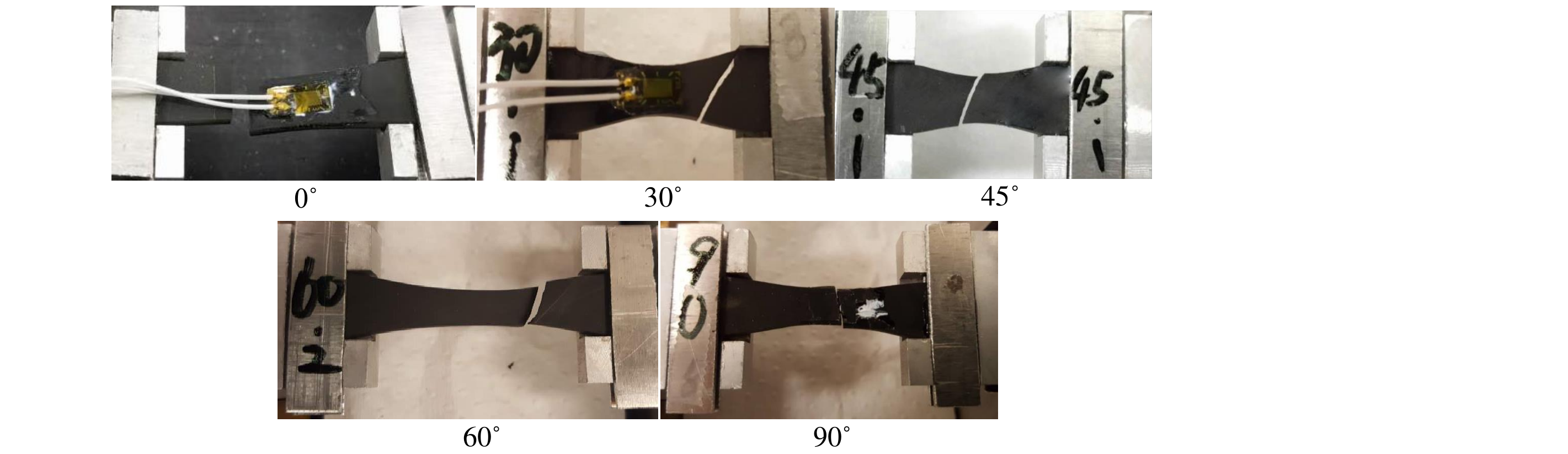}
		\caption{Photograph of specimens after failure in direct tension tests.}
		\label{3.6}
	\end{center}
\end{figure}

Fig.~\ref{3.6} shows the photograph of the failed specimens in the direct tension tests. Similar to the uniaxial compression tests, it can be observed that the failure path was different between specimens with different anisotropy angles. For the specimens with $\theta=0^{\circ}$ and $90^{\circ}$, the failure path was mostly straight and perpendicular to the loading direction, while for the specimens with $30^{\circ} \leq \theta \leq 60^{\circ}$, an acute angle between the failure path and the loading direction was frequently observed. In the case with a lower anisotropy angle ($0^{\circ} \le \theta \leq 45^{\circ}$), it is likely that failure develops along the bedding plane under direct tension; in the case with a steeper angle ($45^{\circ} \leq \theta \le 90^{\circ}$), a more complex interplay between the material matrix and the bedding plane may occur.

\subsection{Brazilian tensile strength}

\begin{figure}
	\begin{center}
		\includegraphics[scale=0.7]{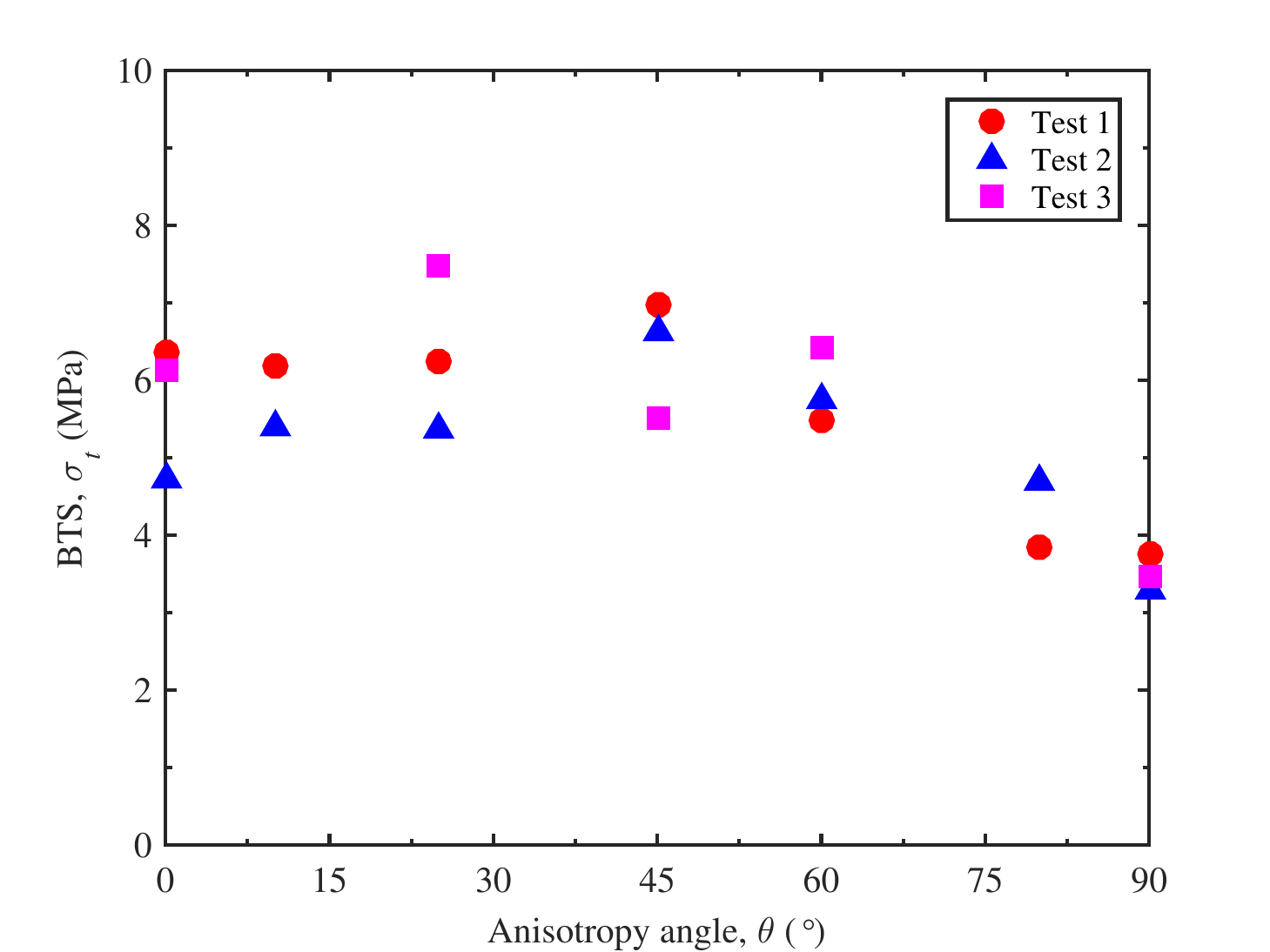}
		\caption{Variation of Brazilian Tensile Strength (BTS) with anisotropy angle.}
		\label{Fig: BTS}
	\end{center}
\end{figure}

\begin{figure}
	\begin{center}
		\includegraphics[width = 0.8\textwidth]{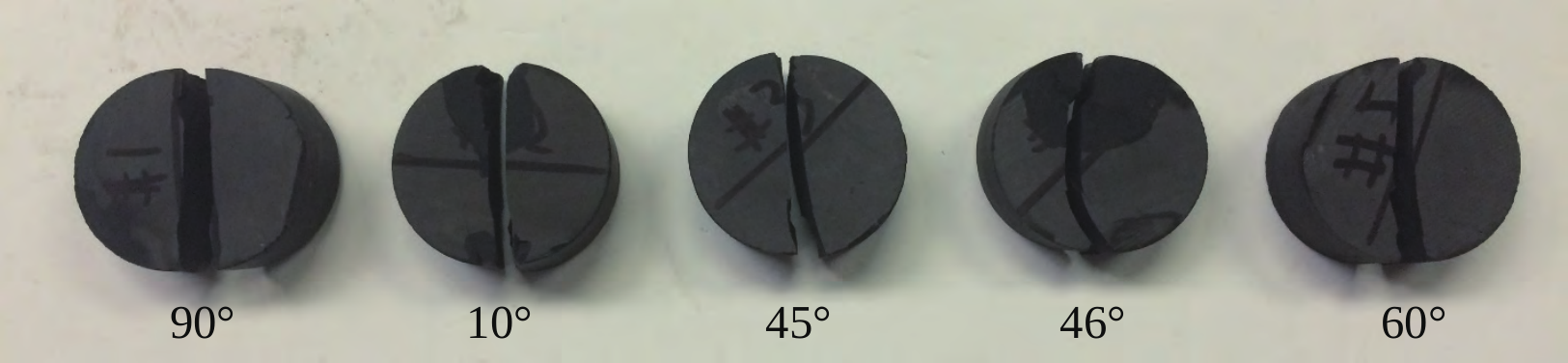}
		\caption{Photograph of specimens after failure in Brazilian tests.}
		\label{Fig: BTSFailure}
	\end{center}
\end{figure}

The results of the Brazilian tests are presented in Fig.~\ref{Fig: BTS}. The value of BTS for each specimen was calculated with the aid of Eq. \ref{Eq:BTS} using the elastic constants obtained from the uniaxial compression tests (Table \ref{tab: Constants}). The obtained BTS values vary from 3.30 to 7.63 MPa, with the maximum occurring at or near $\theta = 25^{\circ}$ and the minimum occurring at $ \theta= 90^{\circ}$. As shown in Fig.~\ref{Fig: BTS}, BTS increases slightly from $\theta=0^{\circ}$ to $25^{\circ}$ and then decreases from $25^{\circ}$ to $90^{\circ}$. The anisotropy ratio calculated by BTS$^{max}/$BTS$^{min}$ is about $2.3$. All investigated specimens were observed to fail suddenly during the tests, and the failure path was found to be mostly along the orientation of the applied loads (loading direction), as shown in Fig.~\ref{Fig: BTSFailure}. However, one may also notice that the failure path occasionally did not intersect with the center point of the disc specimens at $\theta \approx 45^{\circ}$. This is an indication that shear failure may develop along with tensile splitting when the bedding plane is inclined to the loading direction. As a consequence, as also pointed out by Cho et al. \cite{cho2012deformation}, the anisotropic solution for BTS may not be accurate in this case since it does not account for the mixed failure mode.  Indeed, the variation of BTS with $\theta$ that one observes in Fig. \ref{Fig: BTS} can be attributed to different failure modes (splitting, shear, or mixed) occurring at different loading direction with respect to bedding, which has been demonstrated by the numerical analysis in Ref. \cite{li2017multiscale}. 

\subsection{Fracture properties from size effect tests}
The peak loads of the three-point bending tests were recorded and  were used to calculate the nominal strength of each specimen, $\sigma_{Nu}$, which is defined as the maximum tensile stress at failure based on the unnotched cross section, $\sigma_{Nu} = 1.5(S/D)P_u/Dt$, with $P_u$ being the peak load, as reported in Table \ref{tab:fracres}. The apparent fracture toughness, $K_{IcA}$, can then be calculated from $\sigma_{Nu}$ according to classic LEFM, and the apparent fracture energy, $G_{fA}$, can be calculated using Irwin's relation. The obtained $K_{IcA}$ and $G_{fA}$ are also reported in Table \ref{tab:fracres}. It can be seen that there is not only a variation of the calculated fracture properties with the different specimen configurations (due to material anisotropy) but also a variation with specimen size. The size dependency of the laboratory determined fracture properties has been well documented for different kinds of rocks and geomaterials, and is the consequence of material heterogeneity and non-negligible Fracture Process Zone (FPZ). Due to the significant size effect, the fracturing behavior of Marcellus shale in this work cannot be described by means of classic LEFM, calling for nonlinear fracture mechanics of the quasibrittle type. 

\begin{table}[htbp]
	\centering
	\caption{Results of three-point-bending tests on Marcellus shale specimens}
	\footnotesize
	\resizebox{1.0\textwidth}{!}{
	\begin{tabular}{m{1.5cm}>{\centering}m{1.2cm}>{\centering}m{1.2cm}>{\centering}m{1.2cm}>{\centering}m{1.2cm}>{\centering}m{3cm}>{\centering\arraybackslash}m{3cm}}
		\hline
		Type  & Size  & \multicolumn{3}{c}{Nominal Strength, $\sigma_{Nu}$ (MPa)} &  \scriptsize{Apparent Fracture Toughness}, $\bar{K}_{IcA}\pm$SD (MPa$\sqrt{\text{m}}$) &  \scriptsize{Apparent Fracture Energy}, $\bar{G}_f\pm$SD (N/m) \\
		\hline
		\multirow{3}[0]{*}{Arrester} & Large & 6.28  & 5.86  & 5.79  & 0.851$\pm$0.055 & 25.344$\pm$3.250 \\
		& Medium & 7.20  & 6.16  & 9.67  & 0.837$\pm$0.143 & 24.904$\pm$8.671 \\
		& Small & 9.67  & 8.63  & 8.20  & 0.720$\pm$0.093 & 18.292$\pm$4.793 \\
		\multirow{3}[0]{*}{Divider} & Large & 6.70  & 6.64  & 5.80  & 0.967$\pm$0.045 & 24.815$\pm$2.291 \\
		& Medium & 7.84  & 9.35  & 9.19  & 0.852$\pm$0.033 & 19.272$\pm$1.482 \\
		& Small & 9.22  & 9.90  & 8.63  & 0.675$\pm$0.050 & 12.121$\pm$1.810 \\
		\multirow{3}[0]{2cm}{Short-Transverse} & Large & 5.64  & 5.47  & 5.63  & 0.820$\pm$0.043 & 35.913$\pm$3.714 \\
		& Medium & 7.00  & 7.13  & 6.90  & 0.768$\pm$0.010 & 31.486$\pm$0.819 \\
		& Small & 7.92  & 8.11  & 8.94  & 0.642$\pm$0.049 & 22.084$\pm$3.437 \\
		\hline
	\end{tabular}}%
	\label{tab:fracres}%
\end{table}%

In order to obtain the unique fracture properties of the material independent of the specimen and testing method, the Size Effect Law (SEL) proposed by Ba\v zant \cite{bazant1997fracture} was adopted to analyze the experimental data. Details of the analysis can be found in Ref \cite{li2017characterization}. The obtained fracture properties of the material are listed in Table \ref{tab:fracprop}. The size effect method provided not only the fracture toughness, $K_{Ic}$, and fracture energy, $G_{f}$, of the material, but also the effective FPZ length, $c_f$, which is a quantity regarding the material characteristic length used in quasibrittle fracture mechanics. 

\begin{table}[htbp]
	\centering
	\caption{Fracture properties of Marcellus shale from size effect tests}
	\resizebox{0.62\textwidth}{!}{%
    \begin{tabular}{lrrr}
		\hline
		Type  & \multicolumn{1}{l}{$G_f$ (N/m)} & \multicolumn{1}{l}{$c_f$ (mm)} & \multicolumn{1}{l}{$K_{Ic}$ (MPa$\sqrt{\text{m}}$)} \\
		\hline
		Arrester & 29.0  & 0.731 & 0.912 \\
		Divider & 37.9  & 2.99  & 1.20 \\
		Short-Transverse & 44.8  & 1.23  & 0.917 \\
		\hline
	\end{tabular}}%
	\label{tab:fracprop}%
\end{table}%

\section{Discussions }	

\subsection{Comparison of static deformability in tension and compression}\label{discuss_ten_comp}
The assumption of Marcellus shale being a transversely isotropic elastic medium was used in the analysis of the experimental data presented above, and has been validated for the cases of uniaxial loading condition, in both tension and compression. The applicability of this assumption needs further examination when extended to multiaxial or even more complex loading conditions, especially in the presence of both tension and compression. In this section, tensile and compressive deformability are compared. 

\begin{figure}
	\begin{center}
		\includegraphics[scale = 0.7]{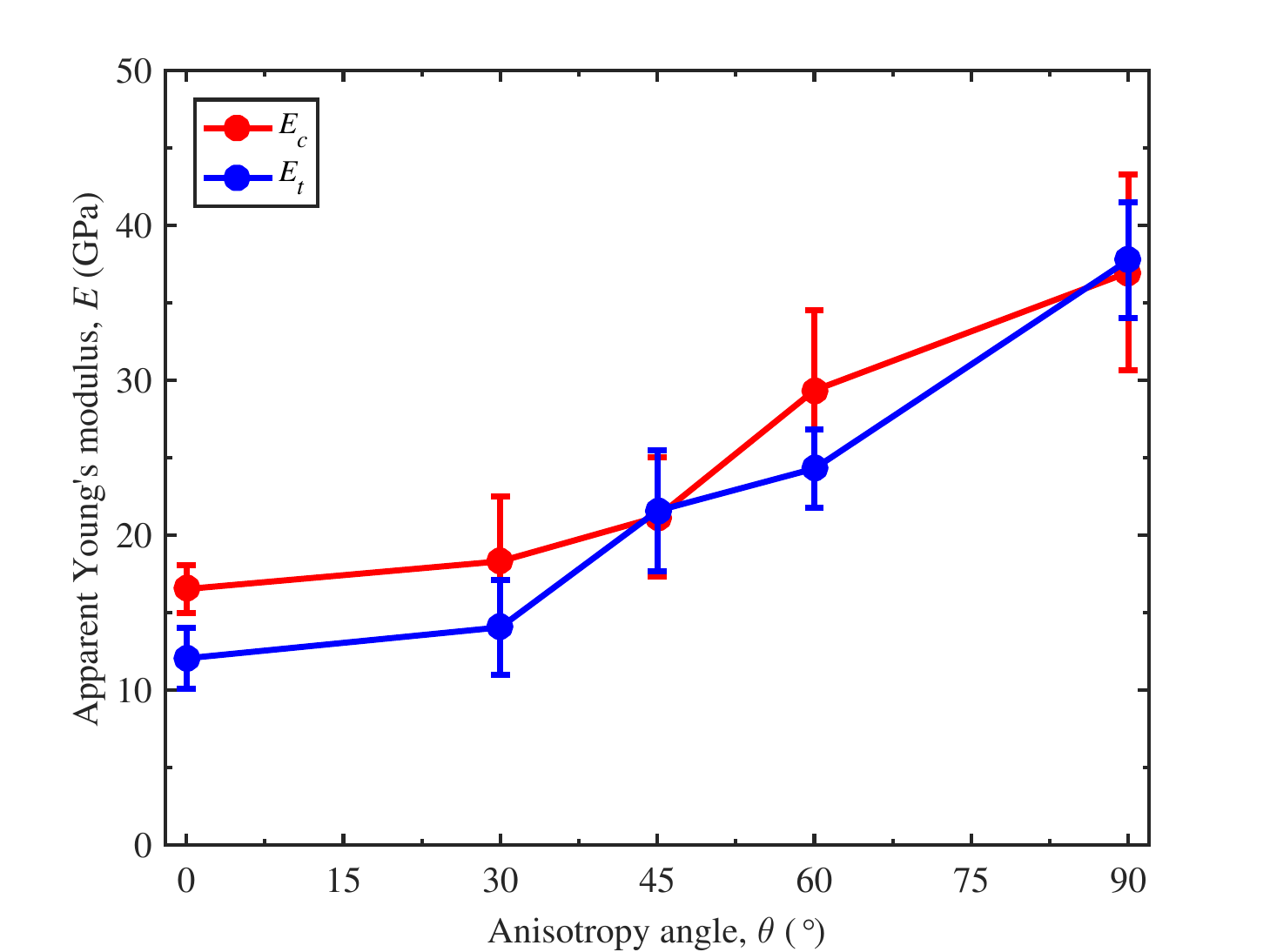}
		\caption{Comparison of compressive moduli and tensile moduli}
		\label{Fig:AppE}
	\end{center}
\end{figure}

Fig.~\ref{Fig:AppE} shows the variation of the measured elastic moduli with anisotropy angle from both the uniaxial compression and the direct tension tests. The error bar represents the standard derivation of the measured Young's moduli at each anisotropy angle. Regardless of specimen anisotropy angle, the average values of the compressive moduli are almost always larger than tensile moduli. The ratio of $E_c$ to $E_t$ varies from $1.00$ to $1.37$. 

It has long been observed that many rocks exhibit different deformational properties when loaded in tension and in compression. Generally, the compressive modulus was found to be larger than tensile modulus \cite{haimson1974stresses, hawkes1973deformation, stimpson1993measurement}, which can be attributed to the presence of natural fractures and micro-cracks, or pores for porous rocks. When loaded in compression, rocks gradually become stiffer with an increase of load due to closure of natural fractures or micro-cracks, leading to an increase of the moduli. In terms of shale rocks, it has been observed by many researchers \cite{gale2014natural,ma2017correlative,josh2012laboratory,josh2012laboratory} that there exist a large amount of natural fractures, micro-cracks, and fissures at the shale grain level, which may account for the difference of Marcellus shale deformability under tension and compression. Further study on imaging of these small scale features, and on the development of mechanics models, is needed to understand and quantify this effect. 

\begin{figure}
	\begin{center}
		\includegraphics[scale = 0.57,  trim=18mm 0mm 0mm 0mm, clip]{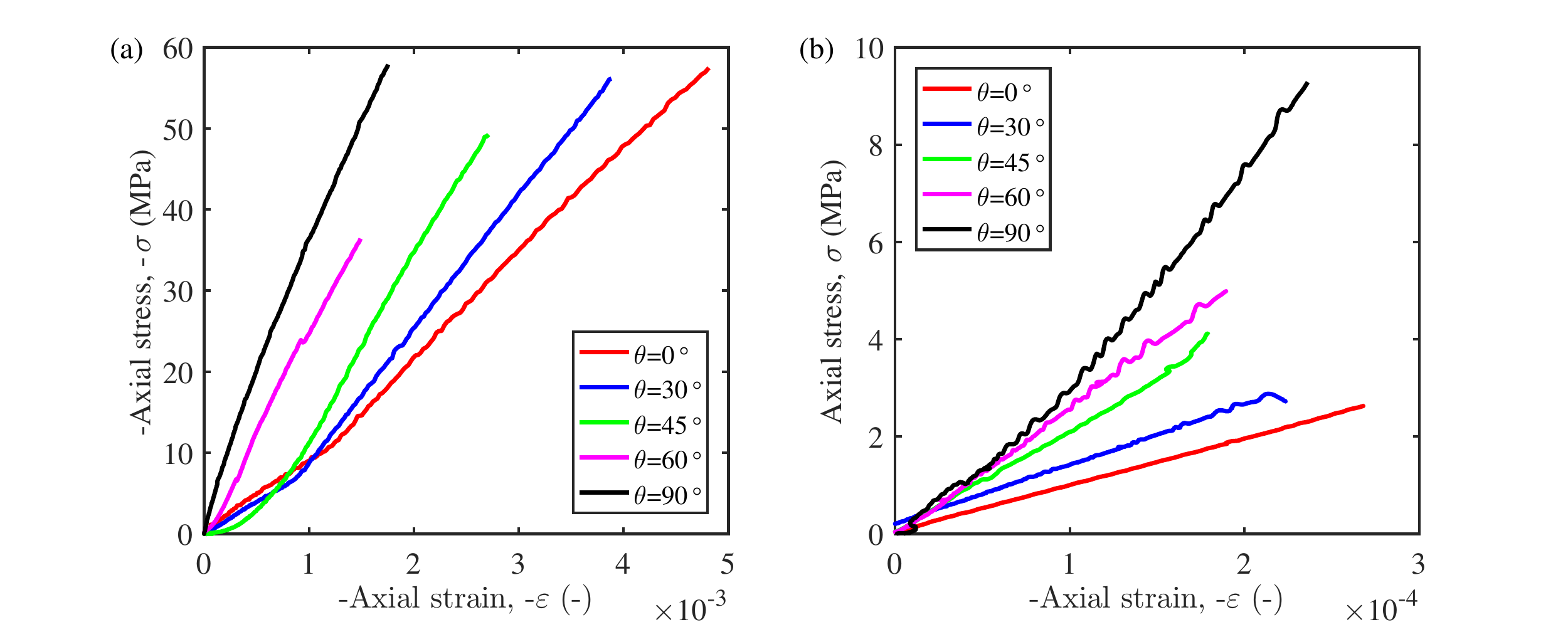}
		\caption{The stress-strain curves for (a) uniaxial compression tests and (b) direct tension tests.}
		\label{Fig:SS}
	\end{center}
\end{figure}

The difference in deformability under compression and tension is also relevant to material anisotropy. Fig.~\ref{Fig:SS}a and b show the typical stress-strain curves for the uniaxial compression and direct tension tests, respectively. It can be seen from Fig. \ref{Fig:SS}a that the relation between axial stress and strain for the uniaxial compression tests was almost linear for most of the specimens with relatively steep anisotropy angle, whereas slight non-linearity can be observed for the specimens with low anisotropy angle at the initial stages of loading. Compared to the curves for the compression tests, the stress-strain relationships for the direct tension tests was mostly linear regardless of specimen bedding plane inclination, as one can observe from Fig.~\ref{Fig:SS}b. 

The observation presented above can be related to the effect of natural fractures and micro-cracks on material deformability under compression. As pointed out by Sayers \cite{sayers2013effect}, cracks in organic-rich shales tend to be bedding-parallel, which may originate from the conversion of kerogen to lower molecular weight hydrocarbons. Therefore, the effect of natural fractures and micro-cracks is maximum when the loading direction is perpendicular to the bedding plane, and minimum when parallel to bedding, which leads to the difference in the degree of nonlinearity for the specimens with different bedding plane orientations, as shown in Fig. \ref{Fig:SS}. This may also explain why the ratio of $E_c$ to $E_t$ at $\theta = 0^\circ$ (loading perpendicular to bedding) is maximum, while the ratio at $\theta = 90^\circ$ (loading parallel to bedding) is minimum (almost 1), as one may note from Fig. \ref{Fig:AppE}. The observation is consistent with the results of $E_v$ and $E_h$ in Table \ref{tab: Constants} as the measured Young's modulus at $\theta = 0^\circ$ is related to $E_v$ and the one at $\theta = 90^\circ$ is related to $E_h$. 

Although the theory of elasticity is generally applicable, this conclusion regarding the difference in deformability in tension and compression ought to be considered when a precise deformation analysis is required in the presence of complex loading conditions. 

\subsection{Comparison of BTS and DTS}
In this work, the tensile strength of the material was measured by both Brazilian and direct tension tests. The strength measurements provided by these two methods, however, show a considerable difference. A comparison of BTS and DTS is provided in this section. 

\begin{figure}
	\begin{center}
		\includegraphics[scale = 0.7]{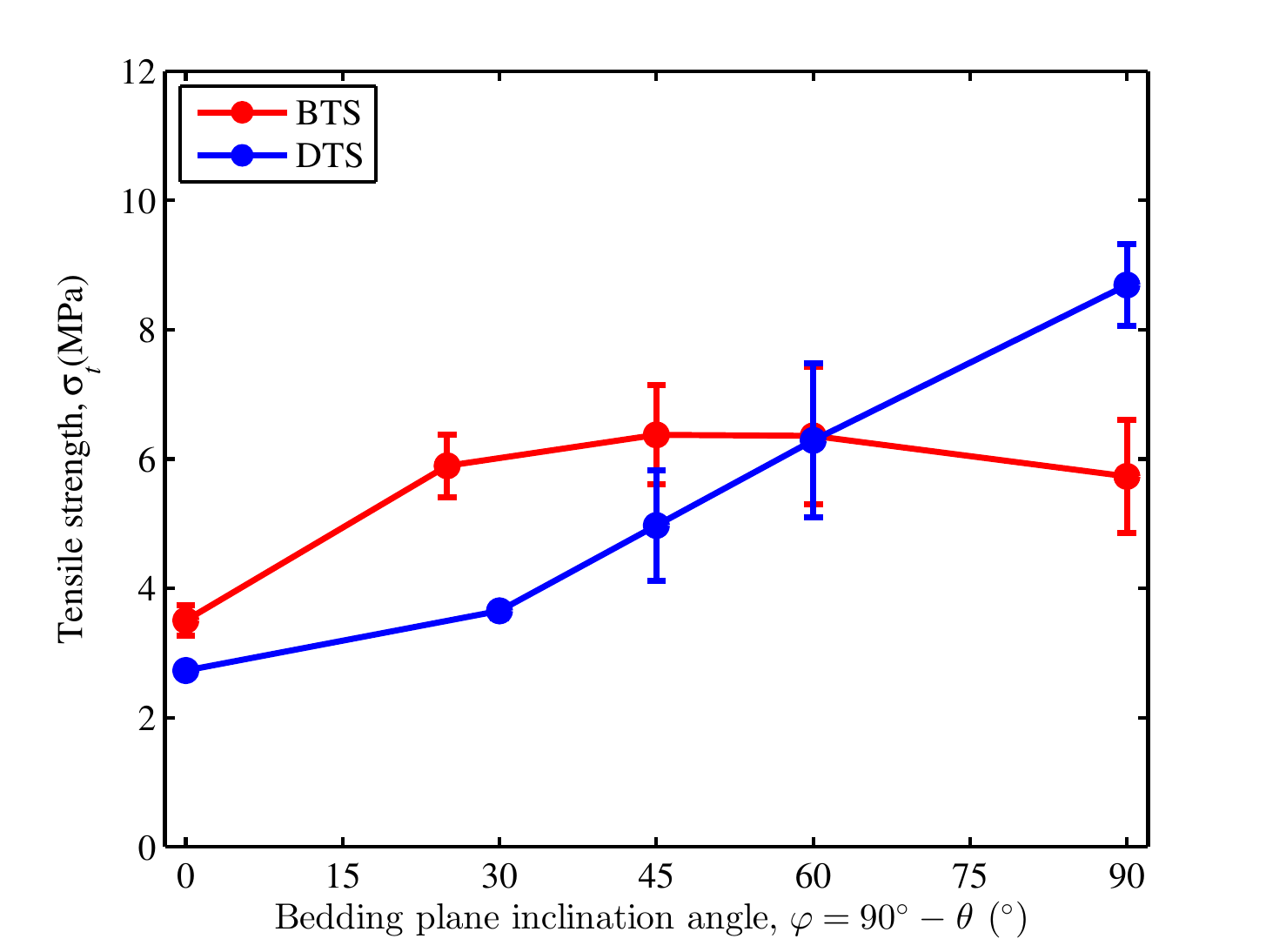}
		\caption{Comparison of Brazilian tensile strength and Direct tensile strength}
		\label{Fig: BTSDTS}
	\end{center}
\end{figure}

The variations of BTS and DTS with anisotropy angle need to be clarified first. In Brazilian tests, the maximum tensile stress is thought to be on a plane perpendicular to the loading direction, whereas in direct tension tests it is uniformly distributed across the specimen cross section and parallel to the loading direction. Hence, BTS at anisotropy angle $\theta$ is comparable to DTS at $90^{\circ} - \theta$. To enable a meaningful comparison between BTS and DTS, the bedding plane inclination angle, denoted by $\varphi$, is defined as the angle between the specimen bedding plane and the plane on which the maximum tensile stress acts. The angle is calculated as $\varphi=90^\circ-\theta$ for Brazilian tests and $\varphi=\theta$ for direct tension tests. Fig.~\ref{Fig: BTSDTS} plots BTS and DTS against $\varphi$. It can be seen that BTS is larger than DTS for the specimens at $0^{\circ} \leq \varphi \leq 45^\circ$, whereas BTS is smaller than DTS at $\varphi = 90^\circ$. 

A large volume of work has been devoted to seeking the relationship between the strength of rocks and the properties one obtains from common testing methods including the Brazilian and direct tension tests, but different conclusions have been drawn. For the Brazilian test, a number of researchers stated that BTS overestimates the tensile strength of rocks \cite{pandey1986deformation,efimov2009rock,fuenkajorn2010laboratory} mainly because specimens are under a biaxial stress state. However, some studies also suggested that BTS may underestimate the tensile strength when the ratio of compressive to tensile strength is low \cite{coviello2005measurement,fairhurst1964validity}. Indeed, calculating the ratio of UCS to DTS based on the results presented in Table \ref{tab:Tension} and \ref{tab:compression}, one can find that this ratio varies with anisotropy angle, and is high at low anisotropy angle (e.g., UCS/DTS = 22.63 at $\theta = 0^\circ$) yet low at high anisotropy angle (e.g., UCS/DTS = 6.60 at $\theta = 90^\circ$). Therefore, one may justify that BTS overestimates the tensile strength of the material at low $\theta$ but underestimates it at high $\theta$. This conclusion is consistent with the observation regarding the difference between BTS and DTS shown in Fig. \ref{Fig: BTSDTS}. Another aspect that may hamper the application of Brazilian test is a lack of a closed-form solution which could account for the mixed failure mode frequently observed for anisotropic rocks. Even for isotropic materials, it has been occasionally observed that failure initiated from the points away from the center of the specimens \cite{fairhurst1964validity,hooper1971failure,hudson1972controlled,swab2011analysis}, which violates the assumptions of analytical solutions. In this case, seeking numerical solutions could be an alternative way of extracting material properties from experimental data. 

The direct methods for measurement of tensile properties have rarely been employed in rock mechanics laboratories mainly because of unavoidable bending, torsion, and anomalous concentrated stress associated with direct tensile loading. Even though well-designed specimen geometries (Fig. \ref{Fig:Specimens}c) and tensile grips (Fig. \ref{Fig:TensileSetup}) were adopted in this work to minimize their effects, it was occasionally observed that failure did not initiate from the center of the waisted specimens. In addition, DTS was defined as the maximum nominal tensile stress applied on the original specimen cross section; however, this does not account for the inclined failure path frequently observed in the tests. Further work is needed to better interpret the results of the direct tension tests. 

Finally, one may note that neither the analysis of the Brazilian tests nor that of direct tension tests accounts for the size effect, which in fact may not be small, as demonstrated by a number of researchers \cite{rocco1999size,thuro2001scale,coviello2005measurement}. Further scaling studies will be needed to tackle the uncertainties of the direct and indirect measurements introduced by the size effect.  

\subsection{Size dependence of apparent fracture toughness}
The fracture tests on the TPB specimens with increasing size show that there exists a significant size dependency of the fracture toughness determined by means of classic LEFM. Indeed, the term \emph{fracture toughness} is widely used in laboratory and field study for characterizing rock fracturability, yet there seems to be some confusion between fracture toughness as a unique material characteristic and an apparent one (i.e., a structural property). To avoid confusion, the apparent fracture toughness measured at specific specimen size and geometry is denoted by $K_{IcA}$, whereas fracture toughness of material is denoted by $K_{Ic}$. 

\begin{figure}
	\begin{center}
		\includegraphics[scale = 0.7]{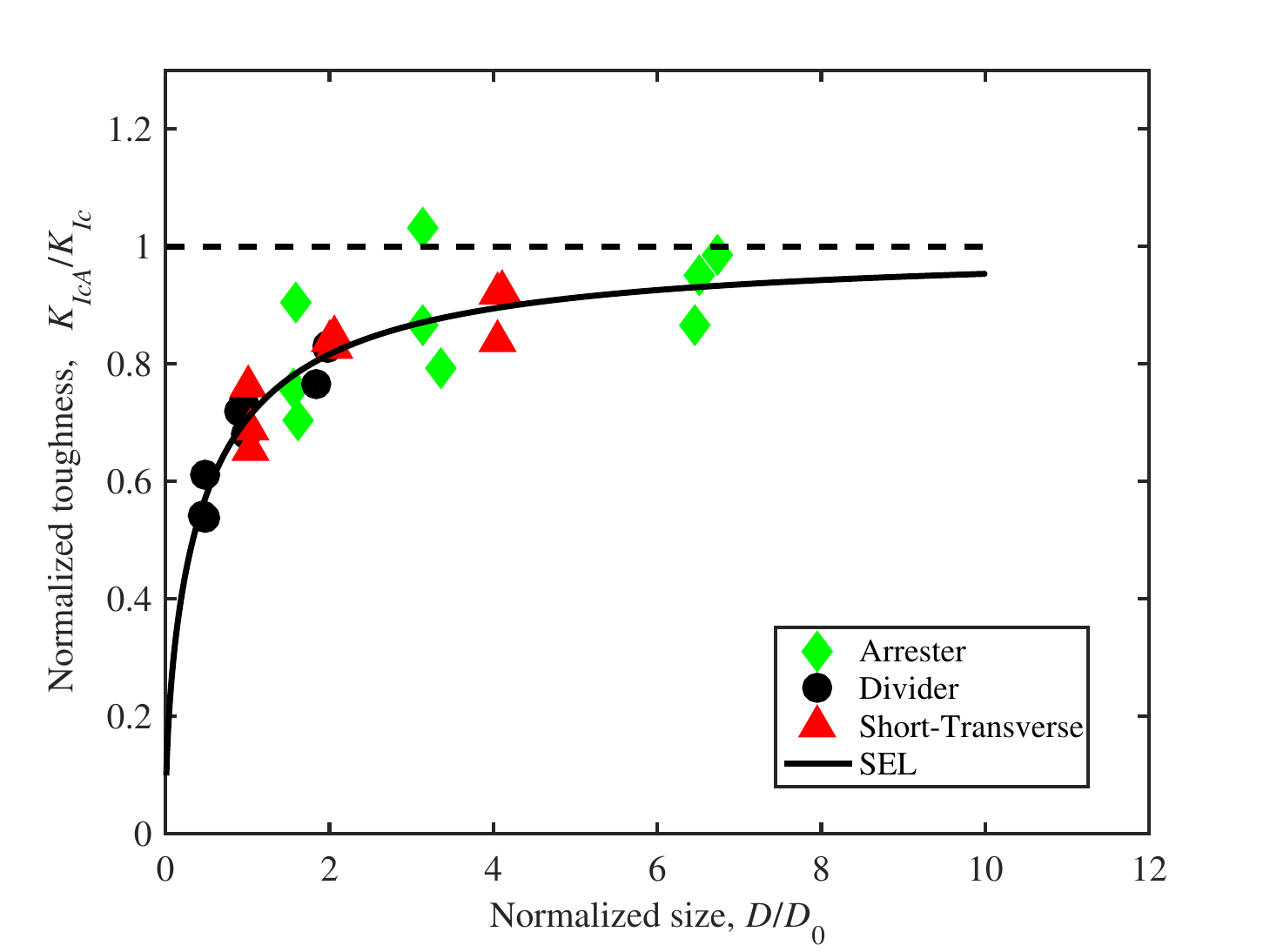}
		\caption{Variation of normalized apparent fracture toughness $K_{IcA}/K_{Ic}$ with normalized size $D/D_0$.}
		\label{Fig:KIc}
	\end{center}
\end{figure}

The apparent fracture toughness was normalized by $K_{Ic}$, and was plotted against the normalized size of the investigated specimens, i.e. $D/D_0$, as shown in Fig. \ref{Fig:KIc}. The variable $D_0$ is called the transitional size which describes the transition from ductile to brittle behavior with increasing structure size, and was calculated by taking into account the FPZ length of the material and the specimen geometry. It can be seen from the figure that the apparent toughness of the specimens increases with the normalized specimen size. 

The black solid line shown in Fig. \ref{Fig:KIc} represents the variation of $K_{IcA}$ predicted by SEL. The predicted trend agrees with the experimental data and shows that $K_{IcA}/K_{Ic}$ gradually increases and eventually converges to the asymptotic value 1 as $D/D_0 \rightarrow \infty$. One can conclude that in order to apply classic LEFM and approximate $K_{IcA}$ by $K_{Ic}$, the specimen size needs to be sufficiently large. In some cases, increasing the specimen size to the required value in order to apply LEFM is not practical. Therefore, the size effect method provides an unique advantage as it does not impose a strict requirement for specimen size and geometry but requires only the knowledge of the peak load from several specimens with increasing size. 

These conclusions ought to be taken into account in various situations relevant to geological engineering design, construction, and operation where a large traction-free crack can grow prior to failure, and when extrapolation from small-scale laboratory tests to field-size structures is needed. In particular, the effect of size becomes extremely important in hydraulic fractures \cite{detournay2016mechanics,chau2016growth,li2017spherocylindrical}.

\section{Conclusions}

Deformability, strength, and fracturing properties of Marcellus shale were studied in this work. The elastic properties and strengths were determined from uniaxial compression, direct tension, and Brazilian tests. In addition, seismic velocities were measured and used to calculate dynamic elastic properties. All of the tests were performed on  specimens with different bedding plane inclinations with respect to loading directions (anisotropy angles) to explore the material anisotropy. Three-point-bending fracture tests were also conducted on specimens with increasing size. Based on the results and analysis, the following conclusions can be drawn:

1)	The five elastic constants based on the transversely isotropic model were calculated by using the least-squares method from the experimental results of the uniaxial compression and direct tension tests, respectively. The obtained shear modulus normal to the plane of isotropy agreed with the value calculated from Saint-Venant's empirical equation. The assumption of transverse isotropy in pure compression and tension conditions was validated by comparing the variation of the apparent Young's modulus with anisotropy angle obtained from the experimental measurements and the theoretical predictions. 

2) The measured P-wave velocities, which varied with anisotropy angle, agree with the ones predicted by the theory of elasticity. The Young's moduli determined dynamically from the seismic velocity measurements were about 50\% larger than the ones from static measurements. 

3) A difference was observed in the elastic properties measured from the uniaxial compression and direct tension tests. The compressive moduli were generally larger than the tensile moduli. The maximum ratio of compressive to tensile moduli occurred at $\theta = 0^\circ$, associated with the slightly nonlinear stress-strain curves in the compression tests, while the minimum ratio occurred at $\theta = 90^\circ$. 

4) The measured Uniaxial Compression Strength (UCS), Direct Tension Strength (DTS), and Brazilian Tensile Strength (BTS) all varied with anisotropy angle, and were related to different failure modes. A pronounced reduction of UCS was found when the shear failure plane was aligned with the inclined bedding plane at $\theta \approx 60^\circ$, supporting the assumption of bedding planes being planes of weakness. DTS increased monotonically with anisotropy angle, while BTS showed a general trend of decreasing with an increase of anisotropy angle.  A curved failure path and an inclined failure plane were observed in the Brazilian and the direct tension tests, respectively, when the loading direction was neither perpendicular nor parallel to bedding. These observations could be attributed to mixed splitting and shear failure modes, but further studies are needed  for a quantitative understanding. 

5) BTS was found to be larger than DTS from $\theta = 0^\circ$ to $45^\circ$, but smaller than BTS  at $\theta =90^\circ$. 

6) The fracture properties calculated from the measured peak loads and by using Linear Elastic Fracture Mechanics (LEFM) were dependent on the specimen size. Theoretically accounting for the size effect provided an indirect way of measuring the unique fracture properties, including fracture energy, toughness, and effective Fracture Process Zone (FPZ) length, independent of testing methods. 

7) In addition to deformation and strength anisotropy, significant anisotropy in the calculated fracture properties was observed. 

\section*{Acknowledgements}
The authors would like to thank Professor B. Sageman for providing the Marcellus shale samples used in this study. Assistance with devices provided by Professor H. Espinosa, Professor G. Buscarnera, Professor J. Qu and their group members is greatly appreciated. This work made use of the Materials Characterization and Imaging Facility and the Center for Sustainable Engineering of Geological and Infrastructure Materials (SEGIM) at Northwestern University.

\section*{References}

\bibliography{shale}

\end{document}